\documentclass[journal]{IEEEtran}
\usepackage{color,graphicx,amsmath,amssymb,amsthm,epsfig,mathrsfs,cite,bm,graphics}
\usepackage{algpseudocode}
\usepackage{algorithmicx}
\usepackage{algorithm} 
\usepackage[utf8]{inputenc}
\usepackage{amsmath}
\usepackage{amsfonts}
\usepackage{amssymb}
\usepackage{placeins}
\usepackage{graphicx}
\usepackage{bbm,bm}
\usepackage{graphicx,graphics,subcaption}
\usepackage{graphicx,graphics}
\usepackage{tikz}
\usepackage{multirow}
\usetikzlibrary{automata, positioning, arrows,shapes.multipart,fit}
\usepackage[normalem]{ulem}
\usepackage{diagbox}
\usepackage{todonotes}
\presetkeys{todonotes}{color=blue!20}{}

\newcommand{\revAdd}[1]{{#1}}

\title{Leveraging online learning for CSS in frugal IoT network}
\author{ Nancy Nayak, Vishnu Raj and Sheetal Kalyani\\
\hspace{-0 cm}Department of Electrical Engineering\\ Indian Institute of Technology Madras, Chennai, India 600 036 \\
\texttt{\{ee17d408@smail,ee14d213@ee,skalyani@ee\}.iitm.ac.in}
}

\begin{document}
	\maketitle
	\begin{abstract}
		We present a novel method for centralized collaborative spectrum sensing for IoT network leveraging cognitive radio network. Based on an online learning framework, we propose an algorithm to efficiently combine the individual sensing results based on the past performance of each detector. Additionally, we show how to utilize the learned normalized weights as a proxy metric of detection accuracy and selectively enable the sensing at detectors. Our results show improved performance in terms of inter-user collision and misdetection. Further, by selectively enabling some of the devices in the network, we propose a strategy to extend the field life of devices without compromising on detection accuracy.

	\end{abstract}
    \section{Introduction}
Internet of Things (IoT), with many interconnected devices of varying traffic and Quality-of-Service (QoS) requirements, is considered as the next technological revolution after the Internet \cite{GUBBI20131645,ATZORI20102787,6710070}. Pre-allocating the limited spectrum appropriately to cater to the needs of thousands of devices with varying requirements and capabilities will be a bottleneck for the proliferation of IoT. To exploit the full potential of IoT, adopting a cognitive radio network (CRN) \cite{788210,7006643} will be essential. The licensed users of the spectrum in a CRN are called primary users or PU. The unlicensed radio users called secondary users or SU, sense and attempt to use idle slots from PUs. Upon finding an idle band, SU acquires that band for transmission while keeping the interference at the minimal level. 

In this work, we assume that the SUs are IoT devices that do not have any time-critical information for transmission. One approach is to treat the problem of spectrum access as a game theory-based online learning problem \cite{8270374}. However, when a very large number of SUs simultaneously try to acquire a free channel for transmission in IoT scenarios, it can result in collision and data loss because of uncoordinated spectrum sensing and acquisition procedure by the SUs \cite{DBLP:journals/corr/abs-1804-11135}. In this work, the focus is on improving the sensing of channel state. Channel allocation after sensing can be handled by various scheduling algorithms \cite{bayhan2013scheduling,zhang2018artificial} and is not the focus of this work.

Several methods have been proposed for robust sensing of the spectrum including energy detectors, waveform based techniques, matched filters, cyclo-stationary based sensing and radio identification based sensing \cite{4796930,1399240}. However, the fading nature of the wireless channel, hidden PUs, shadowing, etc. can degrade the quality of the locally sensed data at each SU. Instead of relying only on the locally sensed observations, information from all the SUs in a CRN can be fused at a central location called fusion center (FC) to identify the state of the channels with high confidence and is called collaborative spectrum sensing (CSS). In CSS, the locally sensed observations from the SUs can be combined in different ways. Several CSS techniques such as OR \cite{1542627}, AND\cite{1542650}, confidence voting (CV)\cite{4446521} operate on locally sensed binary data for obtaining the state of the channels. CSS under the faded channel scenario is also well-studied \cite{4024390,7161310}. However, these works deal with binary feedback (i.e. \textit{busy} or \textit{idle}) from the individual detectors. Another line of work \cite{4686831}, uses soft information from each of the detectors. Based on the Neyman-Pearson (NP) criterion, \cite{4686831} provides an optimal soft combination scheme that maximizes detection probability. In a fast fading environment, \cite{6682626} uses a mixture of gamma distribution based approximation to find a soft decision fusion scheme. 
A robust cooperative spectrum sensing strategy based on false discovery rate (FDR)\cite{benjamini1995controlling} is proposed in \cite{4446520} to maximize bandwidth utilization by controlling the number of misdetection of the occupied bands and thus providing a guarantee for interference control.
However, most of the previously mentioned works make assumptions on the knowledge of the performances of individual detectors in terms of probability of detection $P_d$ and the probability of false alarm $P_{fa}$. 

Use of machine learning techniques in CSS has become increasingly popular and CSS based on support vector machine was introduced in \cite{6635250}. Spectrum sensing by a single SU using a convolutional neural network (CNN) was proposed in \cite{8302117}, and the same for multiple SUs in \cite{8604101}. A recent study \cite{lees2019deep} shows that machine learning techniques can outperform traditional signal detection methods for spectrum sensing. A comprehensive survey of machine learning techniques for spectrum sensing is available in \cite{arjoune2019comprehensive}. \revAdd{Another widely used method for channel access in CRN is Partially Observed Markov Decision Process which is used to determine which channel is free with high probability and hence reduce the search time for detecting a free channel \cite{zhang2010optimal, zhao2007decentralized,raschella2013use}.}

In this paper, we consider dense CRN where IoT devices do not have dedicated spectra and hence rely on opportunistic spectrum access for data transmission. \revAdd{Due to the spatial placements, different devices will perceive the occupancy state of the same channel differently due to the fading and multi-path losses and one might encounter the classic hidden node problem.} This will result in dissimilar detection performance across the devices. In such a dense network, even if all the SUs are not able to satisfy the SNR condition for correct detection for all PUs, we can expect that at least one SU will be able to do so because of the density of the deployment. This makes CSS an attractive solution for spectrum sensing in IoT case. Though the traditional CSS schemes help in improving the detection accuracy in a CRN, they are not designed to handle the cases where a) SU nodes can have widely varying channel conditions b) each SU node can have different detection capability/accuracy due to being from different vendors. \revAdd{ In another line of work \cite{lee2011enhanced}, a system with heterogeneous SUs with different probability of false alarm ($P_{fa}$) is presented and it proposes a method to achieve a system-wide probability of detection ($P_d$) by finding out the optimal number of sensing samples. Once all the SUs sense the channels with the optimal number of samples, the decisions are sent to FC, and OR is performed to protect the PUs from unnecessary interference. However in our work we propose to combine the default information coming from each of the individual SUs based on their past performance.} In such a case, an ideal algorithm will need to weigh the observations from each SUs according to the quality of information and then combine it to arrive at a final decision about the state of each channel. This is not done in any of the above-cited works. Assuming no prior knowledge about the detection performance of individual SUs, we present a novel online learning framework for CSS in a centralized system. To combine the information from all the SUs according to their relative performance in an online fashion, Hedge\cite{freund1997decision} and Perceptron\cite{Freund1999} are utilized. 

\revAdd{We consider an interweave cognitive radio system \cite{goldsmith2009breaking}, where the SUs transmit at the same power level as PUs. This can cause collisions to PU transmission when an SU tries to acquire a channel without sensing it. Once the SUs make independent observations of the channel state, they transmit the data to FC using a perfect feedback channel \cite{singh2015cooperative,cohen2015distributed,sun2016collaborative,so2016group,yao2017cluster}. The FC then combines the information from individual SUs to arrive upon the state of each channel.}

We show that the decision derived using the proposed method helps in reducing interference and increasing the utilization of the idle spectrum in a CSS framework. Further, if exploited intelligently, the centralized solution can also provide energy savings to SUs by removing the sensing phase from the SUs with poor detection performance. Instead of sensing all of the possible bands for transmission, SUs can be directed to sense on the bands for which their detection performance is higher. This can keep the detection performance at the desired level without any degradation while providing energy benefits to the SUs thus making the network energy-frugal in nature. 

The major contributions of this paper are to
\begin{enumerate}
    \item propose an online learning based CSS framework for adaptively combining information from multiple SUs based on their detection performance. This addresses the practical problem that different SU may have different detection capabilities and different fading environments.
    \item present a solution to control the FDR  in the described framework and hence further improve the usage of idle spectrum for transmission.
    \item introduce a mechanism to improve the energy efficiency  and hence lifetime of the IoT devices by selectively disabling the poor performing SUs from sensing the channel with negligible degradation in sensing performance.
    \item \revAdd{present a variation of the proposed method which can handle dynamic CRN environments.}
\end{enumerate}
In summary, we present a novel online learning based CSS technique that adaptively adjusts to the individual performances of SUs.
We present empirical results for the performance of proposed online learning methods and compare them with popular CSS techniques available in the literature across multiple scenarios. We also present a comparison against an offline pre-trained deep learning (DL) method \cite{goodfellow2016deep} and show that the proposed approaches can perform close to that of the deep learning method. The implication of this is discussed in detail in Sec. \ref{sec:DL}.

\subsection{Notations}
Bold face lower-case letters (e.g. $\textbf{x}$) denote column vectors and upper-case letters (e.g. $\textbf{A}$) denote matrices. Script face letters (e.g. $\mathcal{S}$) denotes a set, $|\mathcal{S}|$ denotes the cardinality of the set $\mathcal{S}$. Element at $j^{th}$ row and $i^{th}$ column position in matrix $\textbf{A}$ is denoted by $a_{ji}$. $Q_{\mathcal{D}}$ is the tail distribution function of random variable with distribution $\mathcal{D}$  i.e. $Q_{\mathcal{D}}(x) =\mathbb{P}(X_{\mathcal{D}}>x)$. $\mathbb{I}_{[x]}$ denotes the indicator function of event $x$ happening.

    \section{Channel Sensing in CRN with Online Learning}

We consider a CRN with a set of licensed users (PU) $\mathcal{P}$, $|\mathcal{P}| = P$ and a set of unlicensed channel users (SU) $\mathcal{S}$, $|\mathcal{S}| = S$. In a typical IoT network, we have $P << S$. Also, assuming that each PU has a dedicated channel, we have $P$ channels in the network for opportunistic access. As there are no channels licensed to SUs, they need to use the channels licensed to PUs when the corresponding channel is \textit{idle}. At the beginning of new transmission, the SUs are required to sense each of the known channels. In a CSS framework, all the SUs cooperate in the process of finding idle channels. This is achieved by sending the observation made by each SU to a coordinating central hub, fusion center, which fuses these data to derive meaningful predictions about the channel state.

Let $o_{ji}$ be the observation made by the $i^{th}$ SU about the $j^{th}$ channel state. This information can be either hard (i.e. \textit{busy} or \textit{idle}) or soft (i.e. detected energy) values. Assuming that the SUs sense all the channels for every transmission, the $i^{th}$ SU produces a vector $\textbf{o}_{i} = [o_{1i},o_{2i},\ldots,o_{Pi}]^T$ of length $P$, where each entry corresponds to the observation from each channel. As FC receives the observation vectors from all the SUs, all these individual vectors can be stacked side by side to form a observation matrix $\textbf{O}$ of dimension $P \times S$. We use $o_{ji}(n), \textbf{o}_{i}(n)$ and $\textbf{O}(n)$ to represent the corresponding quantities at a particular time step $n$. FC operates on this individual information and produces decision vector $\textbf{f}(n) \in [0,1]^P$ with the $j^{th}$ element indicating the decision for state of channel $c_j$. Let the true state of channel at time instant $n$ be $\textbf{g}(n)$ where $g_j(n) \in \{0,1\} \: \: \forall c_j \in \mathcal{P}$.

Online learning is a machine learning paradigm when the learning agent tries to learn from the streaming data itself rather than learning from a dataset collected in the past. In the case of an IoT CRN, where the quality of each of the SU devices for sensing is unknown, online learning can be utilized to learn the quality of each of the SU for channel sensing. Once such a qualitative assessment of each SU for each channel is available, we can use it to combine the data from each of the SU to predict the channel occupancy for each PU channels. We now present a generic online learning framework to combine the information from each SUs followed by two specific approaches for quantifying the quality of each SUs in the network.

\subsection{General framework for online learning}
The online learning framework for combining the decisions from individual SUs works in two phases. In the first phase, it combines the information reported by each SU regarding each of the channel states to arrive at a prediction for the actual channel. In the second phase, the online learning framework should update its belief about the quality of each SU based on the actual channel state.

To quantify the quality of each SU for sensing each of the channel states, the proposed online framework maintains a weight vector for each of the available PU channels. Let $\mathbf{w}_j$ be the weight vector of dimension $S$ maintained by the online learning algorithm for channel $c_j$. Based on the observed performance from each of the SU for channel $c_j$, the framework updates the weights and let $\mathbf{w}_j(n)$ represents the weight vector for channel $c_j$ at time instant $n$. Also, let $\textbf{W}(n)$ of dimension $P \times S$ be matrix obtained by concatenating the individual weight vectors $\mathbf{w}_j(n)$. At every instant, the learning framework takes the recent observation matrix $\mathbf{O}(n)$ and weight matrix $\mathbf{W}(n)$ and creates a channel state decision $\mathbf{f}(n)$. Based on the channel state decision, the FC can then perform channel allocations. A transmission can result in either success or failure. If the transmission is successful, we can infer that the PU channel was \textit{idle} otherwise \textit{busy}. Based on this, FC will get the ground truth (GT) $\mathbf{g}(n)$ about the channel states. 
Based on $\textbf{g}(n)$ and the values $\mathbf{W}$ and $\mathbf{O}$, the learning framework will update the weights for each of the SUs and the process repeats. The proposed online learning framework is given in Alg. \ref{alg:generic}.

Assuming that no prior knowledge about the sensing performance of SUs is known, the algorithm starts by initializing all elements in the weight matrix $w_{ji}$ to the same value $w_0$. In case we have some additional information about the sensing performance of each SUs, then we can accordingly incorporate that to the weight initialization. After initialization, the algorithm proceeds with observing $\mathbf{O}$ and computing the channel state decision with a function $\textsc{ComputeDecision}(\cdot,\cdot)$ and updating the weights using function $\textsc{ComputeNewWeights}(\cdot,\cdot,\cdot)$.

\begin{algorithm}
    \caption{Online learning framework for CSS} \label{alg:generic}
    \begin{algorithmic}[1]
        \State Initialize weight matrix $\mathbf{W}$ as $w_{ji}(1) = w_0 \: \forall i,j$
        \For {$n = 1,2,3,\ldots$}
            \For {$s_i \in \mathcal{S}$}
                \State Obtain observation vector $\textbf{o}_i$
            \EndFor
            \State $\mathbf{O} \gets [\mathbf{o}_1, \ldots, \mathbf{o}_S]$.
            \State $\mathbf{f} \gets \textsc{ComputeDecision}(\textbf{W},\textbf{O})$.
            \State Do channel allocation based on the decision $\textbf{f}$.
            \State Obtain ground truth vector $\textbf{g}$.
            \State $\textbf{W} \gets \textsc{ComputeNewWeights}(\textbf{W},\textbf{O},\textbf{g})$
        \EndFor
    \end{algorithmic}
\end{algorithm}

In the following sections, we discuss how to design the functions $\textsc{ComputeDecision}(\cdot,\cdot)$ and $\textsc{ComputeNewWeights}(\cdot,\cdot,\cdot)$. We propose two online learning approaches which can be used in the framework of Alg. \ref{alg:generic}.
    \section{Hedge based online learning} \label{Hedge}
In a cognitive radio network with collaborative spectrum sensing, we can consider each SU as an \textit{expert} and the sensing results reported by each SU for a channel state as an \textit{expert decision}. Under this framework, the process of computing the channel state from individual decisions becomes the well-known \textit{predicting with multiple experts problem}. To dynamically combine the information from multiple experts in an online setting, a decision-theoretic learning framework called Hedge is proposed in \cite{freund1997decision}. Hedge works by assigning normalized weights to all the experts based on their past performances and then uses these weights to combine the decisions from each of the experts to produce the final decision at FC. Hedge calculates the normalized weight matrix $\textbf{P}(n)$ where $p_{ji}(n)$ corresponds to the normalized weight of SU $s_i$ for channel $c_j$ and is computed as
\begin{align}
    p_{ji}(n) = \frac{w_{ji}(n)}{\sum \limits_{i=1}^{S}w_{ji}(n)}. \label{eqn:normalized_weight}
\end{align}
These normalized weights are indicative of the relative performance of how accurately SU $s_i$ senses channel $c_j$. 

The proposed method can work with any channel detection methods and Neyman-Pearson (NP) \cite{Steven} detector is given as an example.
Consider each SU is deployed with NP energy detector as its detection mechanism. An NP detector computes test statistics $e_{ji}(n)$ as energy in the $N$ samples of the received data and compares with a threshold $\eta$ at each time-step $n$ where $e_{ji}(n)$ is given by,
\begin{align}
    e_{ji}(n)= (1/N)\sum_{s=0}^{N-1}x^2[s]>\eta.
\end{align}
Consider $H_0: x(n)=n_0(n)$ and $H_1: x(n)=s(n)+n_0(n)$ where $s(n)$ is transmitted signal and $n_0(n)$ is the noise.  As $e_{ji}(n)$ is sum of square of $N$ IID Gaussian RVs, the detection hypothesis can be written as \cite{Steven},
\begin{align}
    \frac{e_{ji}(n)}{\sigma^2}\sim \chi_N^2 \quad \text{under }H_0 \label{eqn:test_statistics1}\\
    \frac{e_{ji}(n)}{\sigma_s^2+\sigma^2}\sim \chi_N^2 \quad \text{under }H_1 \label{eqn:test_statistics2}
\end{align}
where $\sigma^2$ and $\sigma_s^2$ are noise variance and signal variance respectively. From (\ref{eqn:test_statistics1}), probability of false alarm $P_{fa}$ can be written as,
\begin{align}
    P_{fa} = P\left(\frac{e_{ji}(n)}{\sigma^2}>\frac{\zeta}{\sigma^2}; H_0\right).
\end{align}
According to NP criterion, for a targeted $P_{fa}$, the threshold to detect a channel is calculated as \cite{Steven},
\begin{align}
    \zeta = \sigma^2 . Q^{-1}_{\chi_N^2}(P_{fa})
\end{align}
where $N$ is the number of samples used for energy detector at each SU. Each SU $s_i$ compares the detected energy $e_{ji}(n)$ for each channel $c_j$ with $\zeta$ to find the state of the channel individually. The prediction of SU $s_i$ about the channel $c_j$ at $n^{th}$ time step is $d_{ji}(n)$ given by, 
\begin{align}
    d_{ji}(n) = \mathrm{1}_{[{e}_{ji}(n) \geq \zeta]}
\end{align}
When the binary information $d_{ji} \in \{0,1\}$ from SUs is sent to FC, then each element in observation matrix is given by, $o_{ji}(n)= d_{ji}(n) \quad \forall i,j$ and in case when \textit{soft} information is sent by SUs to FC then each element in observation matrix is $o_{ji}(n)= e_{ji}(n) \in \mathbb{R} \quad \forall i,j$.

The computed normalized weight matrix $\mathbf{P}(n)$ and the obtained observation matrix $\mathbf{O}(n)$ is first combined to create a real value $\tilde{f}_j(n) \in \mathbb{R}$ for the occupancy of channel $c_j$ as
\begin{align}
    \tilde{f}_j(n) = \sum \limits_{i=1}^{S} p_{ji}(n) o_{ji}(n). \label{eqn:hedge_combining}
\end{align}
The combined information $\tilde{f}_j(n)$ is then compared against a threshold $\gamma_j$ to compute final decision, $f_j(n)$, whether the channel is \textit{busy} or \textit{idle}. 
Later in this section (\ref{subsubsec:combining_hard_decisions} and \ref{combining_soft_decision}) we discuss how to compute  $\gamma_j$.

Once the decision $f_j(n)$ is taken, SUs transmit based on $f_j(n)$. If the final decision $f_j(n)$ at FC about the channel $c_j$ is \textit{idle}, the SUs will transmit according to a scheduling strategy at FC. On the other hand, if the state of final decision $f_j(n)$ at FC about the channel $c_j$ is \textit{busy} then SUs don't transmit to avoid collision. Therefore when the SUs don't transmit, it is not possible for FC to know the ground truth (GT). This results in FC being able to observe the ground truth partially and we call this as approximate ground truth (AGT) $\textbf{g}(n) \in \{0,1\}^P$. \revAdd{ Even though AGT is not the true state of a channel, this is FC's best possible understanding of channel state conditions.}


The instantaneous loss for each detector-channel ($s_i,c_j$) pair is then calculated at each SU using the loss function,
\begin{align}
    l_{ji}(n) = \mathcal{L}({d}_{ji}(n),g_j(n)) = |d_{ji}(n)-g_j(n)| \in \{0,1\}. \label{eqn:hedge_loss}
\end{align}

Whenever the algorithm makes a mistake in predicting the channel state, it incurs a loss of $1$, $0$ otherwise. The algorithm then updates the weight of each pair ($i^{th}$ SU $j^{th}$ channel pair) as
\begin{align}
    w_{ji}(n+1) \gets w_{ji}(n) \beta^{l_{ji}(n)}. \label{eqn:hedge_weight_update}
\end{align}
Here, $\beta \in (0,1]$ is the \textit{learning parameter} and decides the rate at which Hedge will adjust the weights based on the observations.  The structure of \textsc{ComputeDecision()} and \textsc{ComputeNewWeights()} under the proposed method is given in Alg. \ref{alg:hedge_css}. Initially for a particular channel, equal weights will be given to all of the SUs but, as time progresses, Hedge will give higher normalized weight to the SUs which shows lower loss (good accuracy) for that given channel. 

\begin{algorithm}[]
\caption{Hedge based decision making}   \label{alg:hedge_css}
\begin{algorithmic}[1]
    \State Initialize $w_0$.
    \State \textbf{Parameters} Set $\beta \in (0,1]$.
    \Function{ComputeDecision}{(\textbf{W},\textbf{O}}
        \For {$j \in \{1,\ldots,P\}$}
            \State Compute $\gamma_j$ using (\ref{eqn:gamma}).
            \For {$s_i \in \mathcal{S}$}
                \State Calculate $p_{ji}$ using (\ref{eqn:normalized_weight})
            \EndFor
            \State Compute $\tilde{f}_j$ as given in (\ref{eqn:hedge_combining}).
            \State $f_j = 1 \:\: \text{if} \:\: \tilde{f}_j \geq \gamma_j \:\: \text{else} \:\: 0$.
        \EndFor
        \State return $\mathbf{f}$
    \EndFunction
    \Function{ComputeNewWeights}{(\textbf{W},\textbf{O},\textbf{g})}
        \For {$j \in \{1,\ldots,P\}$}
            \For {$s_i \in \mathcal{S}$}
                \State Compute $l_{ji}$ using (\ref{eqn:hedge_loss})
                \State Compute new weights as given in (\ref{eqn:hedge_weight_update}).
            \EndFor
        \EndFor
        \State return $\mathbf{W}$
    \EndFunction
\end{algorithmic}
\end{algorithm}


Let $L_{ji}(T) = \sum \limits_{n=1}^{T} l_{ji}(n)$ be the cumulative loss (sum of misdetection and false alarm) suffered by $s_i$ on detecting the state of channel $c_j$ after $T^{th}$ time step. It can be verified from (\ref{eqn:hedge_weight_update}) that weight of $s_i$ for channel $c_j$ at time step $n$ is $w_{ji}(n) = w_{0} \beta^{L_{ji}(n)}$. Since $\beta \leq 1$, an SU which is inaccurate in sensing the channel state will have a low weight. As time progresses, the best SUs for a channel will have higher normalized weights relative to low-performing SUs. This update of weight based on observed loss will boost the best SUs for each channel with time without any prior information. We would like to reiterate that if any prior knowledge is available, it can also be incorporated into the framework through $w_0$ during the weight initialization phase.

Based on the observation $\textbf{O}(n)$ sent to the FC by the detectors, we propose two different ways to combine the information at FC below. When the detectors send only the locally detected binary decision i.e. $o_{ji}(n) \in \{0,1\}$ of the channel state to the FC, we use Hedge \cite{freund1997decision} with \textit{hard combining} (HC) to make the final decisions. On the other hand, if the SUs can send soft information i.e. $o_{ji}(n)$ is detected energy or confidence in prediction, about the channel state, then the proposed method for combining the information in FC is called Hedge \textit{soft combining} (SC).

\subsection{Combining Hard decisions} \label{subsubsec:combining_hard_decisions}
At time instant $n$, the information from SU $s_i$, $\textbf{o}^{i}(n)$, is a vector with entries $0$ or $1$ corresponding to the observed channel states. At FC, the combined expert decision for each channel $j$ at time step $n$ is calculated as (\ref{eqn:hedge_combining}) and is compared against a threshold $\gamma_j=0.5$ to decide the state of the channel. At the initial stage, when all the SUs have equal normalized weight for a particular channel ($\approx \frac{1}{S}$), this strategy will behave similarly to the majority voting strategy\cite{4446521}. The major benefit of this approach is that the SUs need to send only one bit of information (channel state) to the FC, which reduces the overhead for communication between SU and FC. 

\subsection{Combining Soft decisions}
\label{combining_soft_decision}
Instead of sending the hard decision on channel $c_j$, $s_i \in \mathcal{S}$ can send the detected energy of channel $c_j$ at every instant $n$ as a soft decision directly to the FC and let FC can use this data to make a final decision. Even though SC requires higher overhead, the FC can exploit the granularity of the observation to arrive at better decisions. In case of Soft decision combining, $o_{ji}(n)$ denotes detected energy by SU $s_i$ for channel $c_j$. When only noise is present, from (\ref{eqn:test_statistics1}) we have 
$\mathcal{H}_0: o_{ji}(n) = e_{ji}(n) \sim \sigma^2 \chi_N^2$ 
and under the presence of traffic from (\ref{eqn:test_statistics2}) , we get $\mathcal{H}_1: o_{ji}(n)  = e_{ji}(n) \sim (\sigma_{sji}^2 + \sigma^2) \chi_N^2$, where $\sigma_{sji}^2$ is the signal variance between PU $c_j$ and SU $s_i$. Combining this soft information with Hedge, from (\ref{eqn:hedge_combining}), we get
\begin{align}
    \tilde{f}_j(n) = \sum \limits_{i=1}^{S} p_{ji}(n) \eta_{ji}^2 \psi(n), \label{hedge_combining}
\end{align}
where $\psi(n) \sim \chi_N^2$, $\eta_{ji}^2 = \sigma^2$ under $\mathcal{H}_0$ and $\eta_{ji}^2 = \sigma^2 + \sigma_{sji}^2$ under $\mathcal{H}_1$. So $\tilde{f}_j(n)$ follows a weighted sum of $\chi^2$ random variables. Noting that $\chi_N^2$ distribution is a special case of $Gamma(\frac{N}{2}, 2)$ where $Gamma(k,\theta)$ is gamma distribution with shape $k$ and scale $\theta$. Then
\begin{align}
    \tilde{f}_j(n) = \sum \limits_{i=1}^{S} p_{ji}(n) \tilde{\psi}(n),
                 \label{eq:confluent_lauricella}.
\end{align}
where $\tilde{\psi}(n)\sim Gamma \left( \frac{N}{2}, 2 \eta_{ji}^2 \right)$. The weighted sum of gamma random variables follows Confluent Lauricella distribution\cite{6247438}. As the PDF of Confluent Lauricella distribution is difficult to analyze due to the unavailability of closed form expression, the thresholds are computationally expensive to calculate online at every step, we follow a simple approach of approximating the distribution of weighted sum of gamma random variables to another gamma distribution using the popular moment matching technique \cite{8360549}, \cite{6957529}.

Noting that
\begin{align}
    \tilde{f}_j(n) 
        = \sum \limits_{i=1}^{S} p_{ji}(n) \tilde{\psi}(n) 
        = \sum \limits_{i=1}^{S} \nu(n),
\end{align}
where $\nu(n) = \Gamma \left( \frac{N}{2}, 2 p_{ji}(n) \eta_{ji}^2 \right)$. Let $\Gamma(k_j,\theta_j)$ be another gamma distribution which approximates the behaviour of $f_j(n)$. We get the distribution under hypothesis $\mathcal{H}_0$ by considering $\eta_{ji}^2=\sigma^2$. By equating first moment of these distributions, we have
\begin{align}
    k_j \times \theta_j 
        = \sum \limits_{i=1}^{S} \frac{N}{2} \times 2 p_{ji}(n) \sigma^2. \label{eqn:match_m1}
\end{align}
Observing that $\sum \limits_{i=1}^{S} p_i = 1$, we get $k_j \theta_j = N \sigma^2$. Similarly, equating the variance of these two distributions, we get
\begin{align}
    k_j \times \theta_j^2 = \sum \limits_{i=1}^{S} \frac{N}{2} \left( 2 p_{ji} \sigma^2  \right)^2. \label{eqn:match_m2}
\end{align}
Comparing (\ref{eqn:match_m1}) and (\ref{eqn:match_m2}), we get
$ 
    \theta_j = 2 \sigma^2 \sum \limits_{i=1}^{S} p_{ji}^2
$ and $
    k_j = \frac{N}{2 \sum \limits_{i=1}^{S} p_{ji}^2}.
$
Hence, given a probability of false alarm requirement, the threshold for detection at time instant $n$, $\gamma_j(n)$, for channel $c_j$ can be calculated from,
\begin{align}
    \gamma_j = Q_{\Gamma(k_j, \theta_j)}^{-1} (P_{fa}) \label{eqn:gamma}
\end{align}
The state of the channel $c_j(n)$  at $n^{th}$ time step $f_{j}(n)$ is predicted as \textit{busy} or \textit{free} by comparing the combined soft information $\tilde{f}_{j}(n)$ with the threshold $\gamma_j$ i.e. \textit{busy} if $\tilde{f}_{j}(n)>\gamma_j$  according to (\ref{eqn:gamma}).


\subsection{Controlling False Alarms} \label{sec:FDR}

In CSS, the soft information about a channel $c_j$ is sent to the FC from each of the SUs. The soft information from each of the SUs is then combined at FC using (\ref{hedge_combining}) for Hedge. Hence FC has soft information about $P$ number of channels which corresponds to the state of occupancy of those channels. In a CRN, an FC predicting the state of multiple PUs from the detected energies can be formulated as a multiple hypothesis testing problem \cite{shaffer1995multiple} where each test is the prediction of the actual state of the channel $c_j$ by the FC.

Supposing each of the $P$ tests having a probability of false alarm $P_{fa}$, the probability of getting at least one false positive, termed as family wise error rate (FWER), is given by, 
$
    FWER=1-(1-P_{fa})^P.
$
Expanding the second term using binomial expansion we get,
$
    FWER \approx P \times P_{fa}.
$
Controlling FWER at a level $\alpha$ implies each of the $P$ tests should have $P_{fa}=\alpha/P$.

Consider a multiple hypothesis testing problem with $10$ tests and each with $P_{fa} = 0.05$. Then the FWER is $1-(1-0.05)^{10} \approx 0.40$ and it increases with the increase in the number of tests which means the probability of getting false positive increases as we have more channels in the network. If we want to control FWER at a level $\alpha=0.05$ then $P_{fa}$ for each test should be around $1-(1-0.05)^{1/10}=0.005$ which is very conservative because this lowers the statistical detection power which is defined by the probability of correctly declaring $H_1$, i.e. channel is occupied. So the procedure for controlling FWER reduces the probability of getting false positive at the cost of increasing the probability of getting false negative. There are different ways e.g. Bonferroni correction, Sidak correction \cite{abdi2007bonferroni}, Holm–Bonferroni method \cite{abdi2010holm} etc by which statistical detection power can be improved keeping the probability of false positive at a certain level. A well-known procedure that is designed to control the FWER at a certain level $\alpha$ is \textit{False Discovery Rate} (FDR) procedure given by Benjamini and Hochberg (BH) \cite{benjamini1995controlling}. Using the fact that the channel sensing problem in CRN can be posed as a multiple hypothesis problem, FDR procedure can be used by FC to make the final decision about the state of the channel \cite{4446520, 5937265}. FDR is defined as the expected value of the ratio of false discoveries (declaring \textit{busy} when actually it is \textit{idle}) to the total discoveries (declaring \textit{busy}). Suppose there are $P$ hypotheses corresponding to $P$ channels and among which $p_0$ of them are unoccupied. Assuming that $R$ of total $P$ channels are declared as occupied among which $V$ of them are falsely declared as occupied. Then the false discovery rate (FDR) is defined as,
\begin{align}
    FDR = E\left[\frac{V}{R}\right]
\end{align}
Interested readers are referred to  \cite{benjamini1995controlling} for more details on FDR.

Under the special case when the underlying hypothesis for each of the channel is $H_0$ (channel is \textit{idle}), both FWER and FDR are same i.e. 
$
    FDR= FWER = 1-(1-P_{fa})^P.
$
Controlling FDR is less conservative than controlling FWER in the sense that while FWER tries to control the number of false positives in all of the tests that have been performed, FDR controls the number of false discoveries made among the total discoveries. 
Our contribution is in using the the BH procedure of controlling FDR for the proposed online algorithm Hedge. In the following section, we show how to improve the false alarm rate in CRN using the BH procedure in case of soft combining for Hedge.

\subsubsection{Benjamini Hochberg (BH) procedure to reduce false alarm in CRN}
\label{sec:BH}
Considering $\tilde{f}_j(n)$ as the observed combined soft information at FC, the corresponding p-value\footnote{\textit{p-value is the probability of observing a test statistic as extreme or more extreme in the direction of rejection as the observed value\cite{shaffer1995multiple}}.} $P_j$ for channel $c_j$ is found by, 
\begin{align}
     P_{j}= Q_{\Gamma(k_j,\theta_j)}(\tilde{f}_j(n)). \label{eqn:p_val}
\end{align}

If $P_{(1)}\leq P_{(2)}\leq ...\leq P_{(P)}$ are ordered p-values  and $H_{(j)}$ is the null hypothesis corresponding to p-value $P_{(j)}$ $\forall j \in \mathcal{P}$, then Benjamini Hochberg procedure controls FDR at a level $\alpha$ by the following way. Let $k$ be the largest $j$ for which,
\begin{align}
    P_{(j)}\leq \frac{j}{P} \alpha \label{BH_cond}
\end{align}
then reject all $H_{(j)}$ for $j=1,2,...,k$. The work in \cite{benjamini1995controlling} proves that the above procedure controls the FDR at $\alpha$. Instead of comparing weighted sum of individual soft information with a threshold, by BH procedure FC declares the state $f_j(n)$ of the channel $c_j$ as \textit{busy} if (\ref{BH_cond}) is satisfied. Finding the final decision at FC by BH procedure helps to reduce the fraction of missed slots for transmission to a great extent at the cost of the marginal increase in collision. In a dense CRN with the collision level below pre-specified requirements, the BH procedure can reduce missed opportunity for transmission with a negligible increase in the collision. 

\subsubsection{Switch between traditional Hedge algorithm and Hedge with BH} \label{switch_BH_traditional}
The false alarm rate and hence the number of missed transmission opportunities can be controlled by the introduction of BH-procedure in the systems where FC acts on soft information.  However, this comes at the price of an increase in collisions to PUs. As long as the levels of PU collisions are below the required level, the introduction of BH will benefit the system. But in the real systems, we will not have apriori information of the PU collisions and hence may not be able to select the right strategy. Towards this end, we propose a combination of online learning based CSS with and without BH procedure. We introduce a strategy called \textit{Switch-BH} to use the BH procedure to reduce missed slots for transmission as much as possible and also being under a standard collision requirement ($\tau$). The strategy starts by using the the BH procedure for predicting the channel occupancy. At each time step, FC calculates the fraction of slots having collision. Whenever this fraction of collision goes beyond a predefined value ($\tau$), the \textit{Switch-BH} method switches from the BH procedure to the unmodified online algorithm to compute the final decision at FC. With this simple strategy, the FC will be able to benefit from both the strong false alarm controlling capabilities of BH while resorting to the unmodified online framework when the collisions are observed to be high. Note, $\tau$ is a user parameter the system designer has to fix prior according to the requirements.


    \section{Perceptron based online learning}\label{Perceptron}

Given the rising popularity of deep learning for wireless applications \cite{8604101,lees2019deep,8452950}, we explore the simplest single neuron model for online learning namely perceptron for CSS. Then we proceed to briefly discuss deep learning technique and how to exploit it for CSS. We also highlight the pros and cons of using deep learning when compared to the much simpler Hedge and Perceptron based models.

In this section, we present a modified version of perceptron which fits into our need for CSS in a CRN. Perceptron maintains a weight vector $\textbf{w}_j$ of length $S$ for each channel, $c_j$ and the weights are updated based on the past performances of each expert like Hedge. In Perceptron, for each channel, the weights on each SU are initialized to zero. At time instant $n$, observations from $i^{th}$ SU is a vector $\textbf{o}_i(n)$ whose each element denotes the decision on $j^{th}$ channel  being soft decision like detected energy. The combined expert decision on channel $c_j$ at FC is given by,
    \begin{align}
        \tilde{f}_j(n)= \sum \limits_{i=1}^{S} w_{ji}(n) o_{ji}(n). \label{perceptron_combining}
    \end{align}
The combined expert decision is compared with a threshold $\gamma^p_j$ similar to Hedge and a final decision is taken. Thus FC learns the $w_{ji}$s and $\gamma_j^p$ satisfying the equation,
    \begin{align}
        \sum \limits_{i=1}^{S} w_{ji}(n) o_{ji}(n)-\gamma^p_j=0. \label{eq:hyperplane}
    \end{align}
Once the final decision on the channel being \textit{busy} or \textit{idle} is taken, SUs transmit according to a scheduling strategy at FC. Ground truth is observed and the instantaneous losses for detector-channel ($s_i, c_j$) pair are calculated using (\ref{eqn:hedge_loss}). Whenever an expert $s_i$ makes a mistake when the actual channel state $c_j$ is \textit{busy}, the weight of that expert is updated as,
\begin{align}
    w_{ji}(n+1) \gets w_{ji}(n) + \rho \cdot  o_{ji}(n). \label{eq:weight update perceptron pos}
\end{align}
When the actual channel state $c_j$ is \textit{idle} and an expert $s_i$ makes a mistake, the weight of that expert is updated as,
\begin{align}
    w_{ji}(n+1) \gets w_{ji}(n) - \rho \cdot o_{ji}(n). \label{eq:weight update perceptron neg}
\end{align}

The simple perceptron algorithm in the CSS setting learns both weights $w_{ji}(n)$ and the intercept $\gamma_j^p$ of the hyperplane. The updates happen only when the final decision taken by FC is false negative (i.e. declaring \textit{busy} when the true channel state is \textit{idle}) or false positive (i.e. declaring \textit{idle} when the true channel state is \textit{busy}). In a CRN setting, FC can observe the ground truth only by transmitting. When the final decision about a channel is \textit{busy}, FC does not transmit to avoid collision with PUs and other SUs. So the weight update will happen only when the FC decides to transmit and false positive occurs.


As the observations come from $\chi^2_N$ or $\Gamma$ distribution, the combined expert decision $\tilde{f}_j(n) \sim \sum \limits_{i=1}^{S} w_{ji}(n)\Gamma \left( \frac{N}{2}, 2 \eta^2 \right) $ is a weighted sum of $\Gamma$ distributions.
\begin{align}
    \tilde{f}_j(n) \sim \sum \limits_{i=1}^{S} w_{ji}(n) \Gamma \left( \frac{N}{2}, 2 \eta^2 \right). \label{dis:weighted_sum}
\end{align}
\revAdd{However, a closed form expression for the above distribution is not available to the best of our knowledge, neither is a proper approximation. Hence, we propose to use a histogram fitting method and derive the appropriate threshold empirically. 
Though this a computationally costly procedure, we provide the results for perceptron in the Results section as an alternative method of Hedge to do online learning.
Let $\mathcal{H}$ be the normalized histogram of samples drawn from (\ref{dis:weighted_sum}). 
}
For a predefined $P_{fa}$ we can calculate the threshold $\gamma^p_j$ from the $\mathcal{H}$ distribution as given below, 
\begin{align}
    \gamma_j^p = Q_{\mathcal{H}}^{-1}(P_{fa}). \label{eqn:gamma_p}
\end{align}

Our problem setting can be formulated as a perceptron with known bias term $\gamma_j^p$ computed according to (\ref{eqn:gamma_p}). Hence the objective of finding hyperplane (\ref{eq:hyperplane}) can be written as,
\begin{align}
    \sum \limits_{i=1}^{S} w_{ji}(n) o'_{ji}(n)=0 \label{eq:hyperplanemommatch}
\end{align}
where 
\begin{align}
    o'_{ji}(n)= o_{ji}(n)-\frac{\gamma^p_j}{S \times w_{ji}(n)}.\label{eq:observations}
\end{align}

Instead of learning the bias term along with the weights, here we propose to include the bias term in the observations as Eq. (\ref{eq:observations}) and learn only the weights by updating them by the perceptron algorithm. So on false positive, the weights on that expert are updated as,
\begin{align}
    w_{ji}(n+1) \gets w_{ji}(n) + \rho \cdot o'_{ji}(n) \label{eq:weight update perceptron_mom pos}
\end{align}
and on false negative, the weights on that expert is updated as,
\begin{align}
    w_{ji}(n+1) \gets w_{ji}(n) - \rho \cdot o'_{ji}(n). \label{eq:weight update perceptron_mom neg}
\end{align}

At FC the state of channel $c_j$  at $n^{th}$ time step, $f_{j}(n)$ is predicted as \textit{busy} or free by comparing the combined soft information $\tilde{f}_{j}(n)$ with the threshold $\gamma_j^p$ according to (\ref{eqn:gamma_p}). Depending on the final prediction $f_{j}(n)$ by FC, it instructs SUs (either to transmit or not) and the AGT $\textbf{g}(n)$ is known. The procedure of using modified perceptron method is given in Alg. \ref{alg:perceptron_functions}.
The main difference between original perceptron algorithm and the proposed Perceptron with moment matching method is that, the observations are considered as $o'_{ji}(n)$ instead of $o_{ji}(n)$ and the hyperplane to learn is given by (\ref{eq:hyperplanemommatch}). Further only the weights of this hyperplane are learned, not the intercept.

 \begin{algorithm}[]
    \caption{Perceptron based decision making}   \label{alg:perceptron_functions}
    \begin{algorithmic}[1]
        \State Initialize $w_{ji}(1)=0$
        \State \textbf{Parameters} Set $\rho \in (0,1]$
        \Function{ComputeDecision}{(\textbf{W},\textbf{O}}
            \For {$j \in \{1,\ldots,P\}$}
                \State Compute $\gamma^p_j$ using (\ref{eqn:gamma_p}).
                \For {$s_i \in \mathcal{S}$}
                    \State $o'_{ji}(n)= o_{ji}(n)-\frac{\gamma^p_j}{S \times w_{ji}(n)}$
                \EndFor
                \State $\tilde{f}(n) = \sum \limits_{i=1}^{S} w_{ji}(n) o'_{ji}(n)$
                \If {$\tilde{f}(n)$=1} 
                    \State $f_j = 1$
                \Else
                    \State $f_j = 0$
                \EndIf
            \EndFor
            
            \State return $\mathbf{f}$
        \EndFunction
        \Function{ComputeNewWeights}{(\textbf{W},\textbf{O},\textbf{g})}
            \For {$j \in \{1,\ldots,P\}$}
                \For {$s_i \in \mathcal{S}$}
                    \State Use (\ref{eq:weight update perceptron_mom pos}) and (\ref{eq:weight update perceptron_mom neg}) to update the weights. 
                \EndFor
            \EndFor
            \State return $\mathbf{W}$
        \EndFunction
    \end{algorithmic}
\end{algorithm}

The BH procedure as described in Sec. \ref{sec:FDR} can also be applied to Perceptron based online learning to reduce missed opportunities. Based on the observed collision, switching between traditional Perceptron and Perceptron with BH can also be done in a similar fashion as Sec. \ref{switch_BH_traditional}. 

\subsection{Deep Learning based approach} \label{sec:DL}
When multiple perceptrons/neurons are stacked together, both in width and depth, the learning capability of the model increases \cite{cybenko1989approximation}.
Deep learning is the method of using Artificial Neural Network (ANN) with multiple hidden layers to learn the relationship between input and output and has gained significant popularity in wireless applications. \revAdd{The problem of co-operative spectrum sensing has been recently approached from Deep Learning (DL) perspective \cite{8604101, liu2019ensemble}. Even though these approaches improve performance, they come at a cost of high computational complexity. }

\revAdd{
In \cite{8604101}, authors investigate CSS in a CRN using Convolutional Neural Network (CNN) in a multiple-SU case for combining both hard and soft decisions. Using the fact that individual decisions from nearby SUs and adjacent bands are similar due to the spatial and spectral correlation, and the CNNs can exploit this kind of relationships, a CNN is pre-trained with the observations from the SUs and the true state of the frequency bands. 
However, this is considerably different from our setting. \cite{8604101} considers a single PU with multiple bands, and hence a specific SU has similar sensing accuracy for all of the PU bands.  We consider the scenario where PUs are spatially distributed, and hence each SU has different sensing accuracy for each PU. Therefore, in our setting, the PU channel occupancies are not correlated, and the CNN architecture is not useful. We use a fully connected architecture (FullyCon) 
for our problem setting. Once we have enough data, we can train the model offline and then use the trained network to predict the channel occupancy state. 
}

A dense FullyCon network is initially trained with the observations seen from the SUs for each PU (features) and the corresponding actual ground truth of the PU channel state (labels). The input at $n^{th}$ time step to the deep network is a flattened vector of observation matrix $\mathbf{O}(n)$. \revAdd{In \cite{8604101}, the DL model uses softmax at the final layer which gives the probability of the PU being occupied. A cross-entropy loss between the predicted state and the actual state of the PU is minimized during training.} Whereas in our case, the output is a sigmoid layer of dimension $P\times1$ whose $j^{th}$ element denotes the probability of $j^{th}$ channel being occupied. The loss function is the mean square error (MSE) between the output vector of the deep network and the actual GT of the channels.

One of the main challenges of using deep learning-based approach for an online setting is the high computational complexity involved and the fact that only an approximated ground truth can be observed. Also, the convergence of the algorithm to an optimal solution is governed by the loss surface it encounters and the initialization of weights. Further, the output from the deep online method is not amenable to analyze easily and hence false alarm control procedures like BH-procedure are not straight forward to apply. However once trained, deep learning-based models can achieve good accuracy. In the experiments section, we compare our proposed methods against the offline trained approach also.

\revAdd{
\section{Handling non-stationary environment}
In the previous sections we considered only stationary channel condition and we have used Hedge and Perceptron to determine the SUs which detect the channel occupancy states correctly. As these algorithms were designed for stationary channels and experts, these may not be able to give desired results in non-stationary environments. This is because these algorithms compute the weights over all the experts based on past performances by giving equal importance to all observations. However, in a non-stationary environment, an algorithm should be able to \textit{discount} the past observations in favor of more recent observations. We propose to handle this non-stationarity using a modification named \textit{discounting} to the online algorithms. Discounting in Hedge for the non-stationary environment was introduced in dHedge \cite{raj2017aggregating}. The weight update equation for dHedge is given by,
    \begin{equation}
        w_{ji}(n+1) \gets w_{ji}(n)^{\gamma} \beta^{l_{ji}(n)}. \label{eqn:dhedge_weight_update}
    \end{equation}
    where $0 \leq \gamma \leq 1$ is the discounting factor. Hence, dHedge works by reducing the importance of distant past observations using an exponential weighing scheme. When $\gamma=1$, dHedge is equivalent to Hedge.}
    
    \revAdd{We also propose to discount Perceptron to use it in the non-stationary environment and term it as \textit{dPerceptron} in the following text. Whenever an expert $s_i$ makes a mistake given that the actual channel state $c_j$ is \textit{busy}, the weight of that expert is updated as,
    \begin{align}
        w_{ji}(n+1) \gets \gamma w_{ji}(n) + \rho \cdot  o_{ji}(n). \label{eq:dHedge weight update perceptron pos}
    \end{align}
When the actual channel state $c_j$ is \textit{idle} but an expert $s_i$ makes a mistake, the weight of that expert is updated as,
    \begin{align}
        w_{ji}(n+1) \gets \gamma w_{ji}(n) - \rho \cdot o_{ji}(n). \label{eq: dHedge weight update perceptron neg}
    \end{align}
where $\gamma$ is discounting factor.}

\section{Saving energy by selectively deactivating detectors} \label{sec:energy_saving}
One of the major drawbacks of CRN is that the devices have to spend energy during sensing. This energy consumption can easily drain the battery of IoT devices which are not connected to any continuous power supply. One effective solution to save power, and hence improve the field life of the devices, is to limit the sensing operations performed by the SUs especially when it cannot accurately detect the presence of a PU in a channel. The normalized weight $p_{ji}$ in Hedge is a metric about the performance of each detector. As the Hedge algorithm works with normalized weights, we will exploit these weights to identify channels that cannot be accurately sensed by each of the detectors and we can deactivate the detectors selectively for detecting those channels. Note, in previous works \cite{1542627,1542650,4446521,8604101}, there is no metric in the developed algorithms which gives a semantic interpretation of performance. Hence unlike our Hedge algorithm, those algorithms cannot enjoy an additional benefit of energy-saving without any trade-offs.

Let $\textbf{p}_j(n)$ denote the vectors of normalized weights calculated based on (\ref{eqn:normalized_weight}) for channel $c_j$. Initially, all the entries will be of the same value ($= 1/S$) and with time, the detectors with better accuracy will see their normalized weights increasing. Because of the normalization applied while calculating these weights in (\ref{eqn:normalized_weight}), raising the value for a \textit{good} detector will also pull down the value of an \textit{inaccurate} detector for channel $c_j$. Hence, we can utilize these normalized values to inhibit the detectors from sensing the channels for which their detection performance is poor. The network can follow a strategy such that at any time instant $n$, when the normalized weight of active channel-detector pair falls below predefined threshold $\mu$, FC can command the detector to stop sensing that channel. This will especially be useful when we have a few \textit{good} detectors for a channel and all the other detectors are \textit{inaccurate}; in such a situation, the proposed method can save a considerable amount of energy by shutting down a large number of \textit{inaccurate} detectors from sensing the channel without compromising on the quality of sensing. On the other hand, if all the detectors are performing equally bad for a specific channel, then FC will ideally need more data to derive accurate predictions and all detectors should sense the spectrum. During the online weight update phase, Hedge provides almost equal normalized weights to similar performing devices and hence we can keep all the detectors active in such challenging situations. Because the weights of the perceptron algorithm are not normalized, it is not straight forward to extend this technique for the same. Neither can such energy-saving be achieved in the DL approach.
    \section{Experimental Results}
\label{sec:experimental result}
In this section, numerical results are presented to evaluate the performance of the proposed methods. To show the utility of the proposed approach over the existing methods, we provide comparisons with well-established CSS techniques like OR\cite{1542627}, AND\cite{1542650} and confidence voting\cite{4446521}. In a CRN scenario, there can be multiple reasons for SUs to be at different signal conditions, for example fading, multi-path, different distance of SUs from the PU, etc. In the following study, we consider three different signal condition scenarios for SUs. We provide a comparison of collisions incurred by SU and PU and the missed transmission opportunities (misdetections) by SU.

\subsection{Simulation setup}
We assume that the PUs and SUs are distributed uniformly in a square area. Because of spatial placements, each SU observes different signal condition for each channel (PU). A set of 10 PUs, i.e $10$ channels, are present in the CRN.
Each PU transmits data with power $PT_j = 0 dB$. The observed power of channel  $c_j$ at SU $s_i$ is modeled as
$
    \mathcal{R}_{ij}= PT_j - PL_{ij}, 
$
where all the terms are in dB. $PL_{ij}$ is the pathloss between $j^{th}$ PU and $i^{th}$ SU with a distance $r_{ij}$ and is computed according to Winner II Model \cite{7069496} for CRN, given by
$    PL_{ij} = 20\log(r_{ij})+46.4+20\log\left(\frac{f_c}{5}\right)$. Further, during the transmission, we consider a random packet loss with probability $p_{l} = 0.05$. The PU traffic is modeled using the Hyper-exponential distribution (HED) as suggested in \cite{5506438}. Both ON time and OFF time of PU are modelled using an $M$ component HED random variable $X$ as
\begin{align*}
    f_X^{HED}= \sum_{k=1}^M p_k f_{Y_k}(x),
\end{align*}
where each $Y_k$ is exponentially distributed with rate $\lambda_k$, and $p_k$ is the weight given to $k^{th}$ component with $\sum \limits_{k=1}^{M}p_k = 1$. The simulation parameters used are given in Table \ref{tab:my_label1}. 


\begin{table}[ht]
\begin{center}
    \begin{tabular}{| l | c | l | c |}
        \hline
         Parameter & Value & Parameter & Value \\ 
        \hline\hline
        Number of PUs & $10$    & $P_{fa}$ & $0.05$ \\
        \hline
        Working frequency of PUs & $6$ GHz & Packet loss & 0.05 \\ 
        \hline
        PU Transmit Power & $0$ dB  & Hedge HC: $\beta$ & 0.88 \\
        \hline
        No.of HED components & $3$ & Hedge SC: $\beta$ & 0.99 \\
        \hline
        $\lambda$ & $(0,500]$ & Perceptron: $\rho$ & 0.80 \\
        \hline
    \end{tabular}
    \caption{Parameter used for simulation}
    \label{tab:my_label1}
\end{center}
\end{table}

In a smaller area, all SUs will be able to sense the channel occupancy accurately most of the time because of their proximity to PUs. However, as area increases, the distance between PUs and SUs will also increase and because of the path loss, their detection performance reduces. 
To get different signal conditions at SU, we model the CRN in three different configurations as follows.
\begin{enumerate}
    \item \textit{Good signal condition} (GSC): An area of $1\times1$ $km^2$ with $10$ SUs. Here all the SUs experience good SNR. Approximately $78\%$ of the SUs have probability of correct detection $P_d>0.95$.
    \item \textit{Medium signal condition} (MSC): An area of $8\times8$ $km^2$ with $50$ SUs. Some of the SUs are in good SNR regime while and some are in low SNR condition. Approximately $55\%$ of the SUs have $P_d>0.95$.
    \item \textit{Bad signal condition} (BSC): An area of $8\times8$ $km^2$ with $10$ SUs. All SUs experience low SNR. Just about $1\%$ of the SUs have $P_d>0.95$.
\end{enumerate}

As the channel occupancy sensing mechanism, we employ energy detectors in each SU and to find the threshold for energy detection, we assume a target $P_{fa} = 0.05$. Because of the different signal conditions, the detection quality of each SU in the CRN will be different.



\ifCLASSOPTIONtwocolumn
\begin{figure*}[!t]
    \centering
    \begin{subfigure}{.33\textwidth}
        \includegraphics[width=\linewidth]{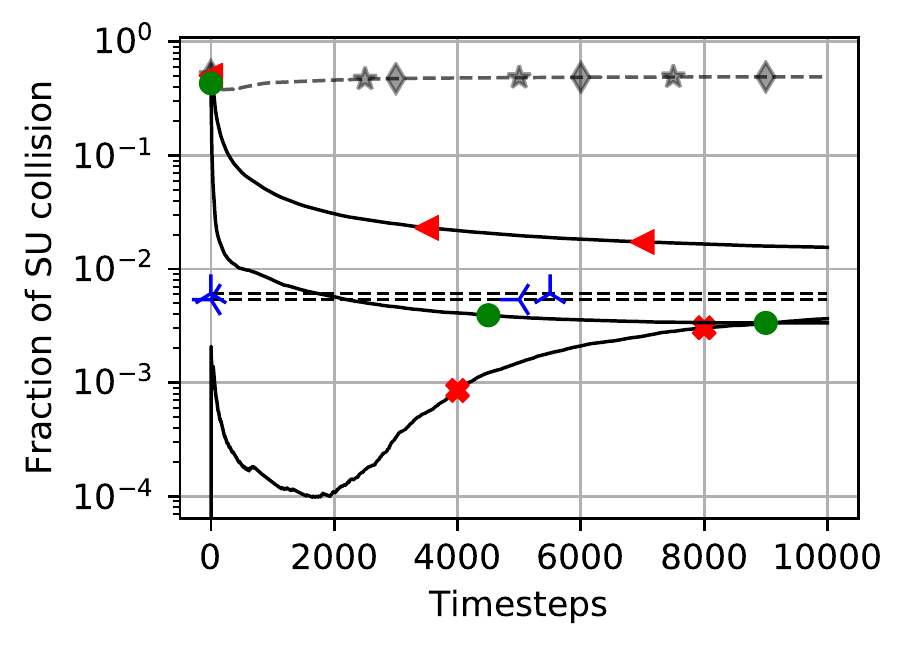}
        \caption{Fraction of SU packet collision in MSC}
        \label{fig:50su_pktcoll_MSC}
    \end{subfigure}%
    \begin{subfigure}{.33\textwidth}
        \includegraphics[width=\linewidth]{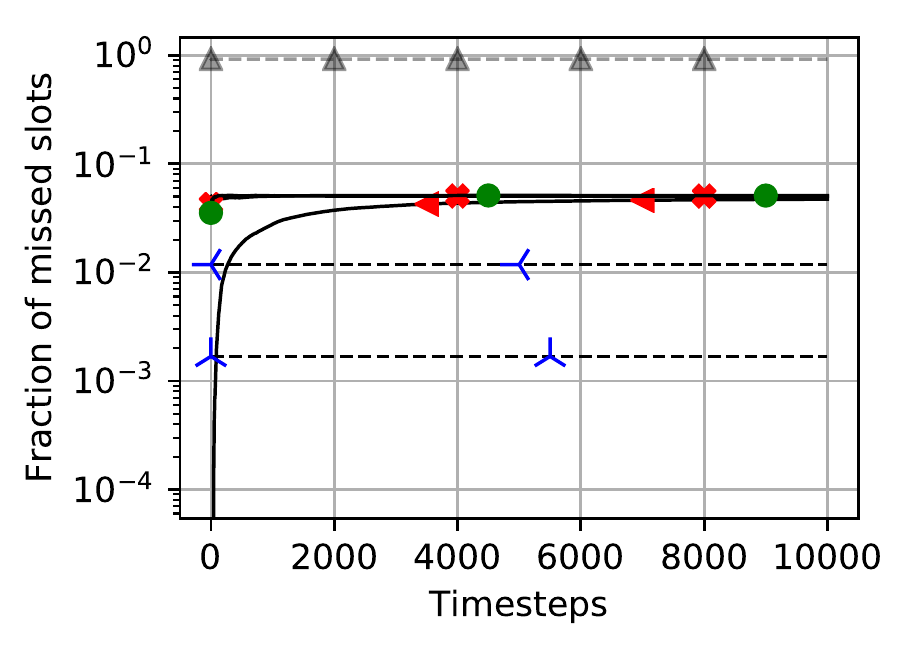}
        \caption{Fraction of missed idle slots in MSC}
        \label{fig:50su_missed_MSC}
    \end{subfigure}%
    \begin{subfigure}{.33\textwidth}
        \includegraphics[width=\linewidth]{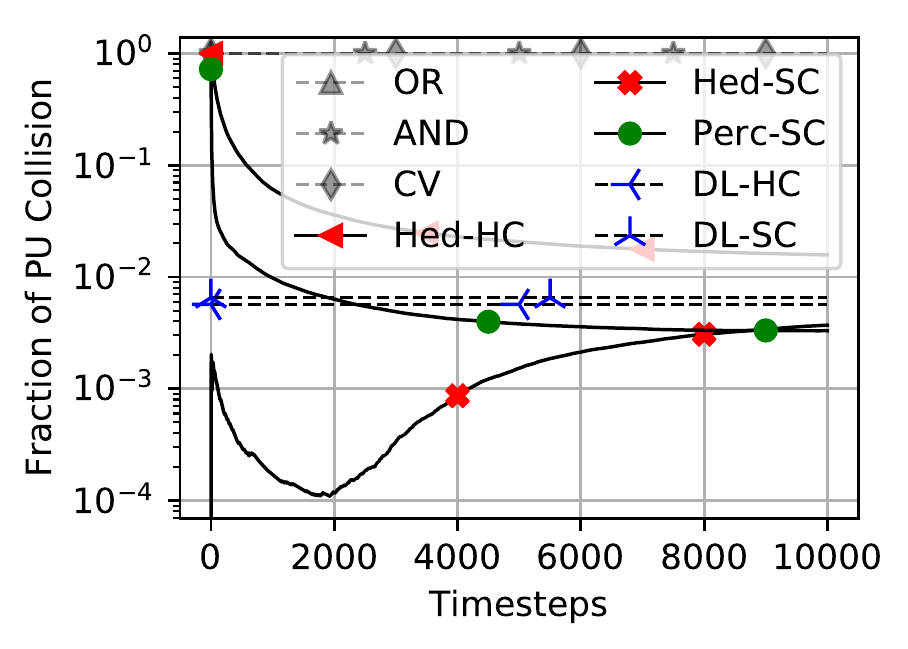}
        \caption{Observed interference at PU in MSC}
        \label{fig:50su_int_MSC}
    \end{subfigure}
    \begin{subfigure}{.33\textwidth}
        \includegraphics[width=\linewidth]{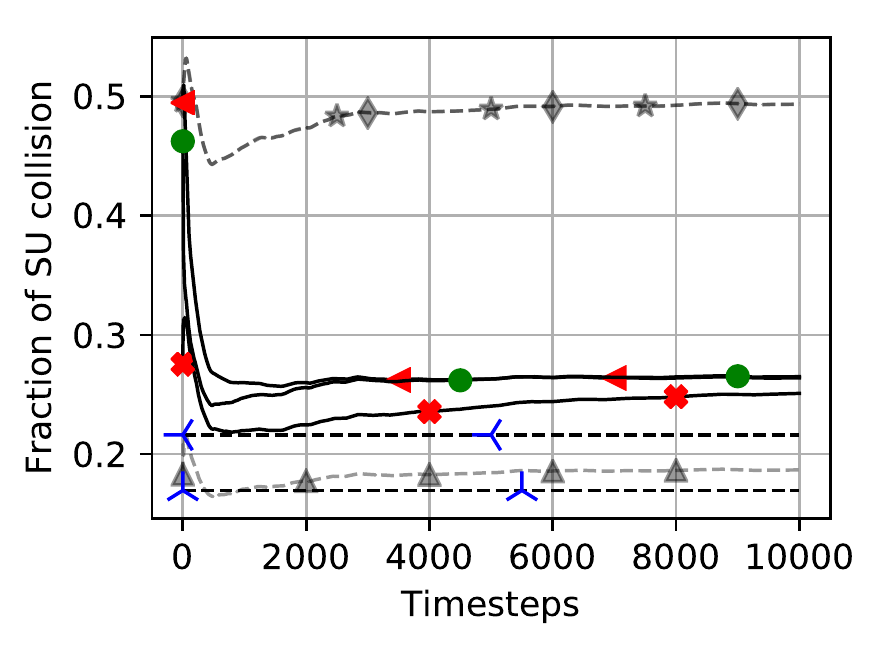}
        \caption{Fraction of SU packet collision in BSC}
        \label{fig:10su_pktcoll_BSC}
    \end{subfigure}%
    \begin{subfigure}{.33\textwidth}
        \includegraphics[width=\linewidth]{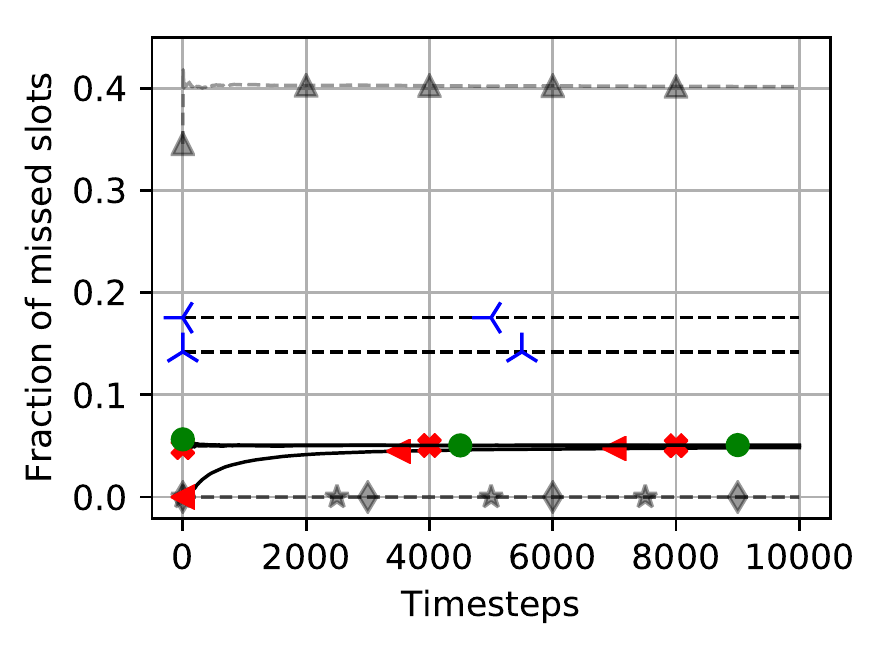}
        \caption{Fraction of missed idle slots in BSC}
        \label{fig:10su_missed_BSC}
    \end{subfigure}%
    \begin{subfigure}{.33\textwidth}
        \includegraphics[width=\linewidth]{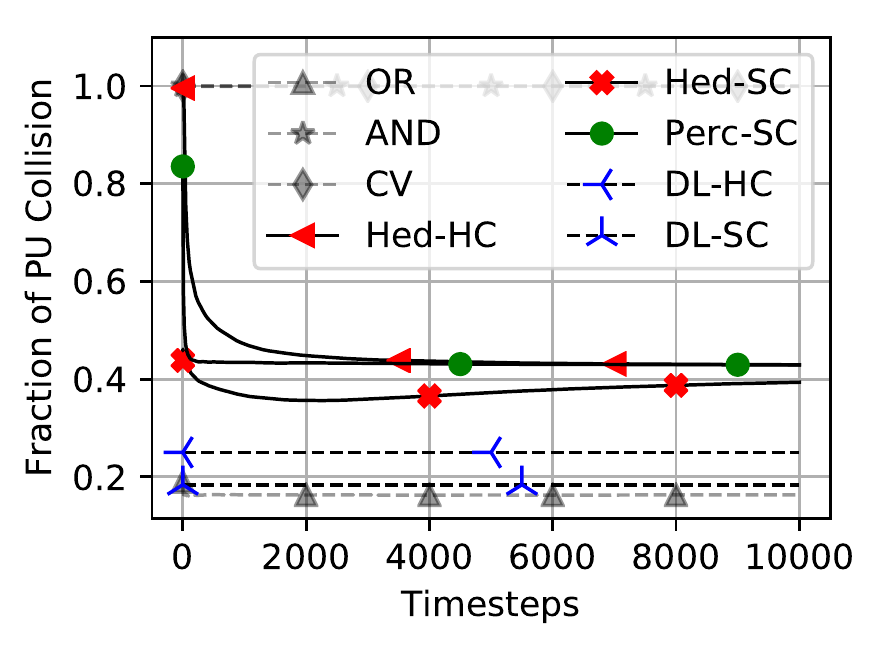}
        \caption{Observed interference at PU in BSC}
        \label{fig:10su_int_BSC}
    \end{subfigure}%
    \caption{Comparison of proposed Hedge-HC, Hedge-SC and Perceptron-SC with traditional OR, AND and CV for MSC and BSC (Plots include the values at every timestep, but markers are put sparsely to improve visibility).}
    \label{fig:Comp_with_traditional}
\end{figure*}
\fi

\begin{figure}[ht]
        \centering
        \includegraphics[scale=0.55]{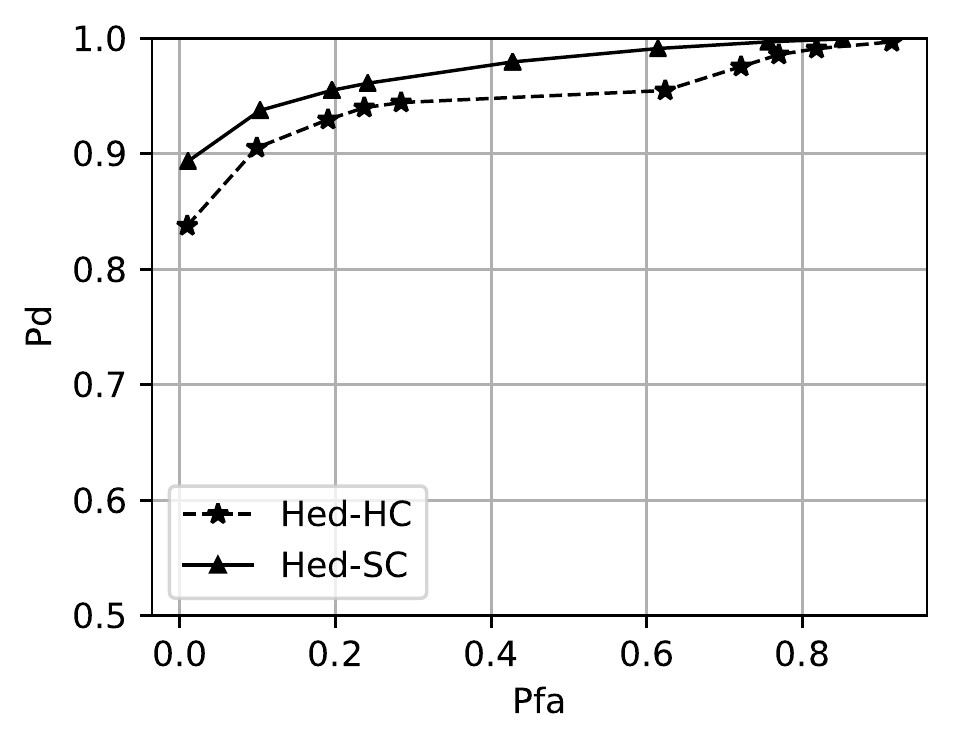}
        \caption{ROC}
        \label{fig5}
\end{figure}

For Hedge-HC at FC, a threshold of $\gamma_j = 0.50$ is used and the decision rule is to declare the channel as \textit{idle} when $f_j(n) < 0.50$, otherwise \textit{busy}. In Hedge-SC and in Perceptron-SC the thresholds $\gamma_j$ and $\gamma_j^p$ are calculated according to (\ref{eqn:gamma}) and (\ref{eqn:gamma_p}) respectively. An SU is allowed by FC to transmit using $c_j$ if the final decision at FC, $f_j(n)$ is \textit{idle} and not to transmit when $f_j(n)$ is \textit{busy}. But initially, the decisions at FC may not be correct because of wrong predictions by the SUs which are in bad signal condition. A wrong final decision $f_j(n)$ at FC can cause either collision or missed transmission opportunity. To find about the quality of predictions from each SU, FC needs feedback about the occupancy state of the channel (Ground Truth). In the next round, this information helps FC to give more weights to those SUs which predicted correctly in the current time-step. FC can obtain feedback about the occupancy state of each of the channels from the success/ failure of SUs' data transmission. When SUs attempt for data transmission two cases can arise. 

\begin{enumerate}
    \item If the final decision $f_j(n)$ is \textit{idle} an SU transmits and can observe the Ground Truth $g_j(n)$ for channel $c_j$ based on whether the transmission was successful or not. If the transmission is successful SU incurs no loss.
    \item If the final decision for $c_j$ is \textit{busy}, SUs are not supposed to transmit. If the SUs are not transmitting on some channel $c_j$, FC doesn't have a mechanism to obtain the ground truth.
\end{enumerate}
Under Case 2, FC cannot observe the actual state of the channel and it proceeds with considering the ground truth of channel $c_j$ as \textit{busy}. This is the AGT observation strategy used in our experiments.



In the two online learning algorithms (Alg. \ref{alg:hedge_css} and Alg. \ref{alg:perceptron_functions}) discussed, weights are updated only when $f_j(n)$ is \textit{idle}. Over time as the SUs with high accuracy gain higher weights, the probability of FC making an error decreases. 

The proposed online learning methods Hedge and Perceptron are also compared with the deep learning approach. The parameters used to train the deep network are fixed after extensive parameter tuning to get the best possible results and are given in Table \ref{tab:my_label3}. Once the network is trained, it is tested with a fresh data set and the average performance of the deep network is compared with other methods.


\ifCLASSOPTIONonecolumn
\begin{table}[ht]
\begin{center}
    \begin{tabular}{| l | c | l | c |}
        \hline
         Parameter & Value & Parameter & Value\\ 
        \hline\hline
        Number of hidden layers & $3$ & Number of neurons in each hidden layer & $2 \times P \times S$ \\
        \hline
        Activation & $tanh$  & Optimizer & SGD-Optimizer\\
        \hline
        Learning rate of batch wise GD & $0.001$ & Batch size & $20$ \\
        \hline
    \end{tabular}
    \caption{Parameters for offline training of  deep network}
    \label{tab:my_label3}
\end{center}
\end{table}
\else
\begin{table}[ht]
\begin{center}
    \begin{tabular}{| l | c |}
        \hline
         Parameter & Value \\ 
        \hline\hline
        Number of hidden layers & $3$ \\
        \hline
        Number of neurons in each hidden layer & $2PS$ \\
        \hline
        Activation & tanh \\
        \hline
        Learning rate of batch wise GD & $0.001$ \\
        \hline
        Batch size & $20$ \\
        \hline
    \end{tabular}
    \caption{Parameters for offline training of  deep network}
    \label{tab:my_label3}
\end{center}
\end{table}
\fi

\ifCLASSOPTIONtwocolumn
\begin{figure*}[ht]
    \centering
    \begin{subfigure}{.33\textwidth}
        \includegraphics[width=\linewidth]{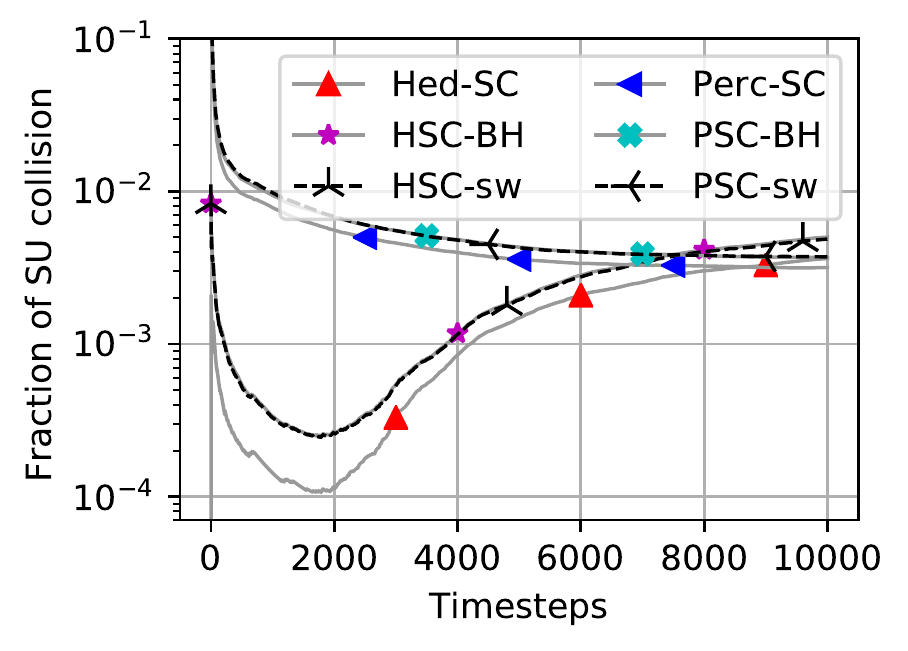}
        \caption{Fraction of SU packet collision for MSC}
        \label{fig:50su_pktcoll_SW_MSC}
    \end{subfigure}%
    \begin{subfigure}{.33\textwidth}
        \includegraphics[width=\linewidth]{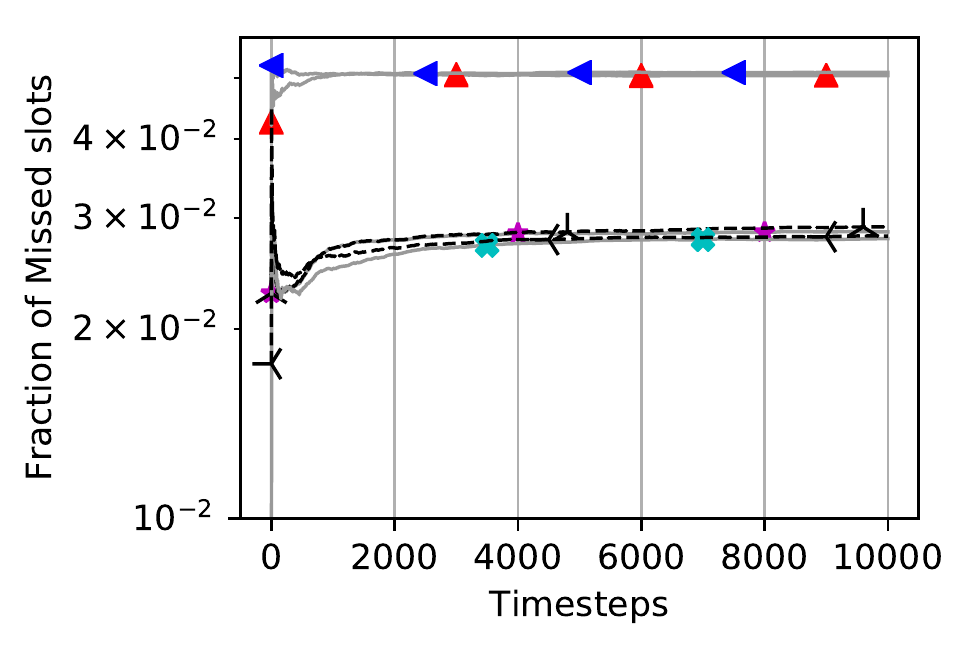}
        \caption{Fraction of missed idle slots for MSC}
        \label{fig:50su_missed_SW_MSC}
    \end{subfigure}%
    \begin{subfigure}{.33\textwidth}
        \includegraphics[width=\linewidth]{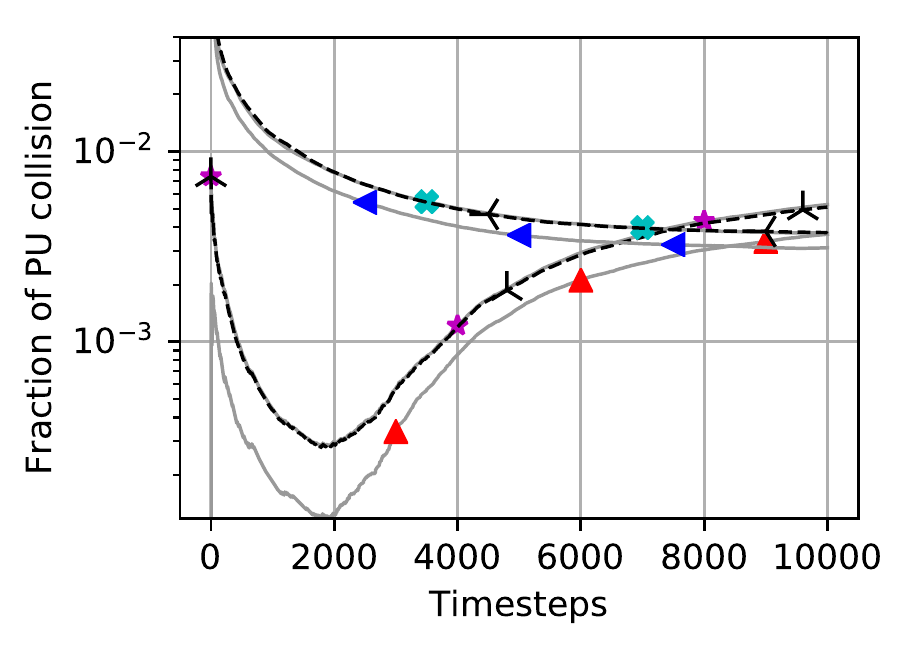}
        \caption{Observed interference at PU for MSC}
        \label{fig:50_su_int_SW_MSC}
    \end{subfigure}
    
    \begin{subfigure}{.33\textwidth}
        \includegraphics[width=\linewidth]{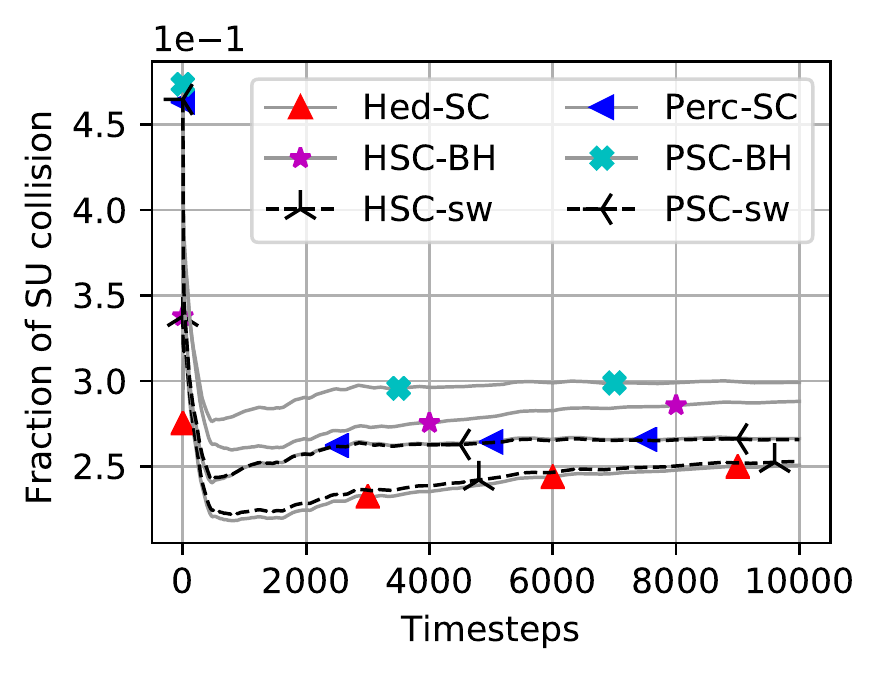}
        \caption{Fraction of SU packet collision for BSC}
        \label{fig:10su_pktcoll_SW_BSC}
    \end{subfigure}%
    \begin{subfigure}{.33\textwidth}
        \includegraphics[width=\linewidth]{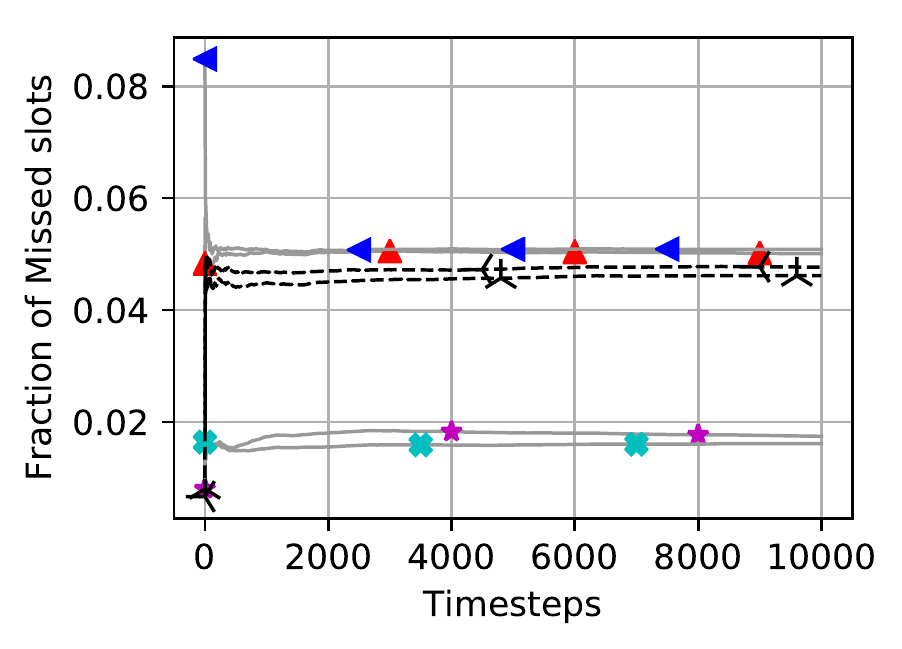}
        \caption{Fraction of missed idle slots for BSC}
        \label{fig:10su_missed_SW_BSC}
    \end{subfigure}%
    \begin{subfigure}{.33\textwidth}
        \includegraphics[width=\linewidth]{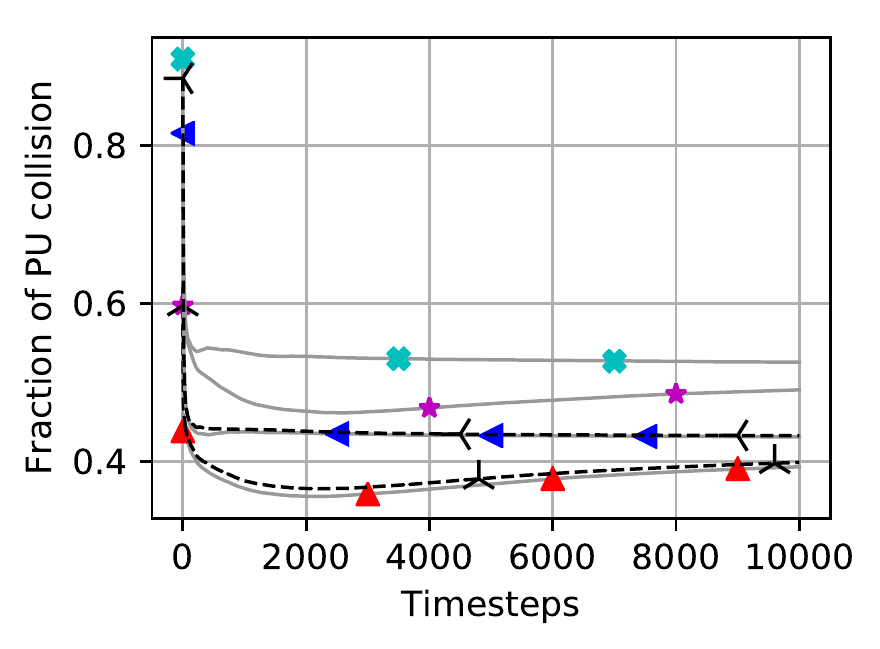}
        \caption{Observed interference at PU for BSC}
        \label{fig:10su_int_SW_BSC}
    \end{subfigure}
    \caption{Comparison of switch strategy in different signal conditions}
    \label{fig:switch}
\end{figure*}
\fi


Assuming the SUs always have data to transmit and will utilize an empty slot in the spectrum if found, we use the following metrics to compare the performances of various approaches against our proposed method. All the metrics are computed for $10000$ time-steps.

\begin{enumerate}
    \item \textit{Fraction of PU collision}: This denotes the interference created at PU due to the collisions. 
    The fraction of PU collision at time step $N$ is given by,
\ifCLASSOPTIONonecolumn
    \begin{align*}
        \text{Fraction of PU collision} = \frac{\sum \limits_{n=1}^N \sum \limits_{j} \mathbb{I}_{[\text{collision observed in }c_j\text{ at }n]}}{ \sum \limits_{n=1}^N \sum \limits_{j} \mathbb{I}_{[c_j\text{ is busy at }n]}}. 
    \end{align*}
\else
    \begin{align*}
        \text{Frac. of PU coll.} = \frac{\sum \limits_{n=1}^N \sum \limits_{j} \mathbb{I}_{[\text{collision observed in }c_j\text{ at }n]}}{ \sum \limits_{n=1}^N \sum \limits_{j} \mathbb{I}_{[c_j\text{ is busy at }n]}}. 
    \end{align*}
\fi
    \item \textit{Fraction of SU collision}: When the PU channel $c_j$ is busy but the final decision is \textit{idle} then SU initiate a transmission and experience collision.
    Then,
\ifCLASSOPTIONonecolumn
    \begin{align*}
        \text{Fraction of SU collision} = \frac{\sum \limits_{n=1}^N \sum \limits_{s \in \mathcal{S}} \mathbb{I}_{[\text{$s$ incured a collision at }n]}}{ \sum \limits_{n=1}^N \sum \limits_{s \in \mathcal{S}} \mathbb{I}_{[\text{$s$ attempts a transmission at }n]}}.
    \end{align*}
\else
    \begin{align*}
        \text{Frac. of SU coll.} = \frac{\sum \limits_{n=1}^N \sum \limits_{s \in \mathcal{S}} \mathbb{I}_{[\text{$s$ incured a collision at }n]}}{ \sum \limits_{n=1}^N \sum \limits_{s \in \mathcal{S}} \mathbb{I}_{[\text{$s$ attempts a transmission at }n]}}.
    \end{align*}
\fi
    \item \textit{Fraction of missed slots}: When the actual state of PU channel is \textit{idle} but the final decision at FC is \textit{busy}, then SU misses a chance to transmit. Let $m_j$ be the number of time-steps channel $c_j$ was free but predicted as busy and $t_j$ be the number of time-steps channel $c_j$ was free. Then,
\ifCLASSOPTIONonecolumn
    \begin{equation*}
         \text{Fraction of missed slots} = \frac{\sum \limits_{n=1}^N \sum \limits_{j} \mathbb{I}_{[\text{Incurred a false alarm at $c_j$}]}}{ \sum \limits_{n=1}^N \sum \limits_{j} \mathbb{I}_{[c_j\text{ is idle at }n]}}.
    \end{equation*}
\else
    \begin{equation*}
         \text{Frac. of missed slots} = \frac{\sum \limits_{n=1}^N \sum \limits_{j} \mathbb{I}_{[\text{Incurred a false alarm at $c_j$}]}}{ \sum \limits_{n=1}^N \sum \limits_{j} \mathbb{I}_{[c_j\text{ is idle at }n]}}.
    \end{equation*}
\fi
    \item \textit{Number of sensing}: The number of sensing done averaged over all the SUs in the network. When SUs are shutdown from sensing channels, this comes down and can be seen as an indicator of energy spent in sensing. Initially, for a particular channel, all of the SUs sense that channel but over time the proposed method can shut down the poor performing SUs for that specific channels and hence the total number of sensing decreases. 
\end{enumerate}

\subsection{Results and discussion}
\label{sec:results}

\setcounter{figure}{5}
\ifCLASSOPTIONtwocolumn
\begin{figure*}[!ht]
    \centering
    \begin{subfigure}{.33\textwidth}
        \includegraphics[width=\linewidth]{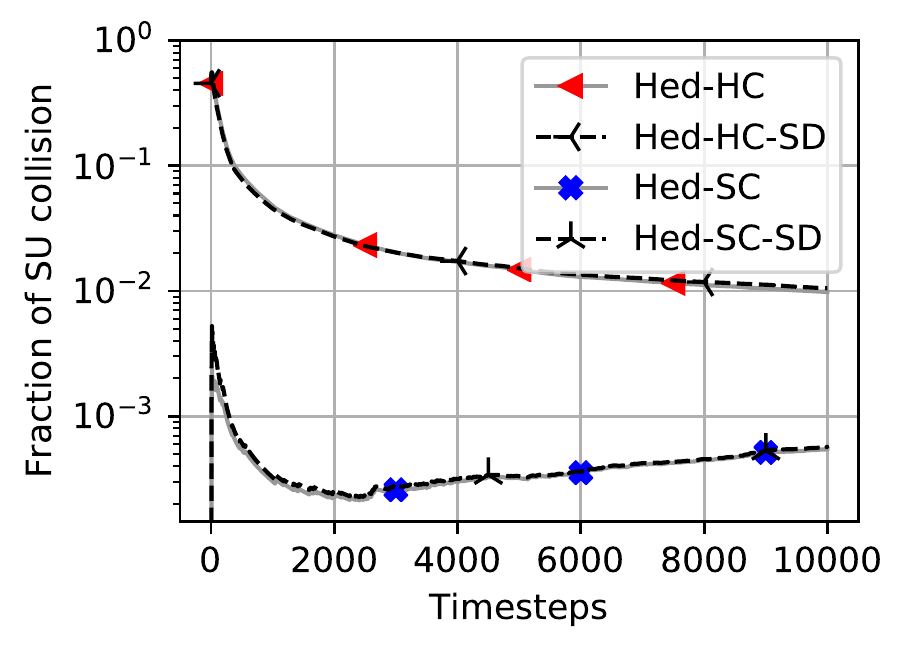}
        \caption{Fraction of SU packet collision}
        \label{fig:50su_pktcoll}
    \end{subfigure}%
    \begin{subfigure}{.33\textwidth}
        \includegraphics[width=\linewidth]{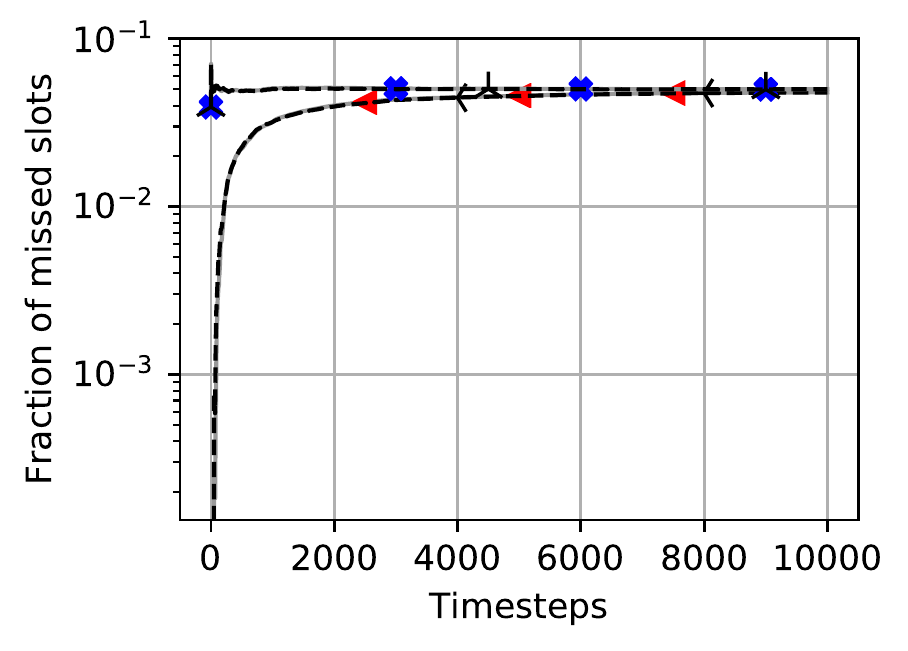}
        \caption{Fraction of missed idle slots}
        \label{fig:50su_missed}
    \end{subfigure}%
    \begin{subfigure}{.33\textwidth}
        \includegraphics[width=\linewidth]{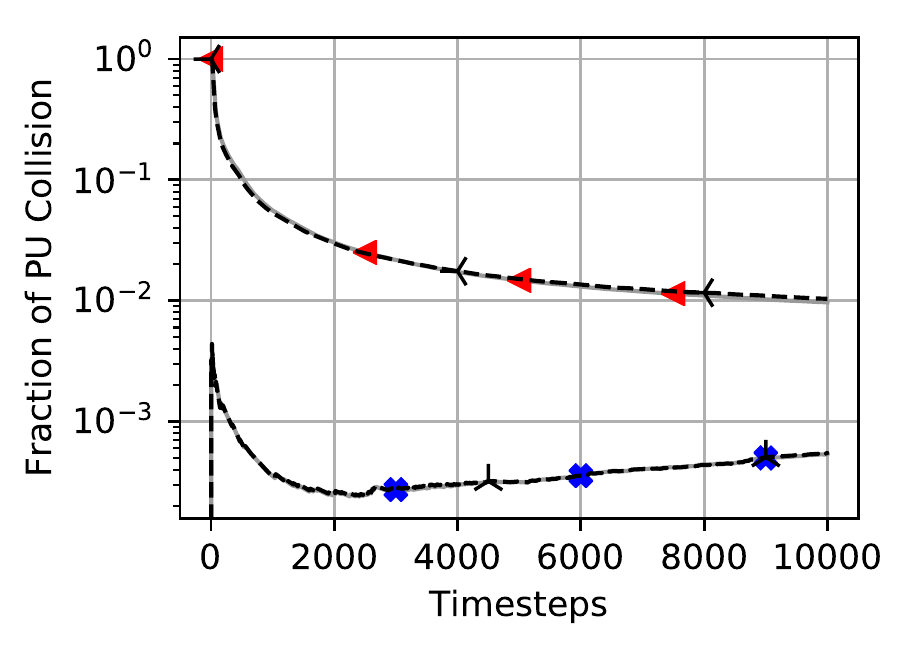}
        \caption{Observed interference at PU}
        \label{fig:50su_int_SD}
    \end{subfigure}%
    \caption{Comparison of metrics with selective deactivation of poor-performing detectors in MSC}
    \label{fig:50SU_result_SD}
\end{figure*}
\fi

A comparison of above-mentioned metrics is given in Fig. \ref{fig:Comp_with_traditional}. The proposed online algorithms Hedge (Alg. \ref{alg:hedge_css}) with hard information (HC) and soft information (SC) combining are denoted by label \textit{"Hed-HC"} and \textit{"Hed-SC"} respectively. Perceptron algorithm (Alg. \ref{alg:perceptron_functions}) is labeled as \textit{"Perc-SC"}. The performance of deep network pre-trained with hard information is labeled by \textit{"DL-HC"} and the same with soft information is labeled by \textit{"DL-SC"}. In this section, we provide results for MSC and BSC. Kindly refer to the appendix for results in GSC.

From Fig. \ref{fig:Comp_with_traditional} we can observe that \textit{Hed-HC} performs better than other traditional CSS techniques \textit{OR}\cite{1542627}, \textit{AND}\cite{1542650} and Confidence Voting (\textit{CV})\cite{4446521} in terms of SU packet collision and interference at PU, keeping missed opportunity at a low level. \textit{Hed-SC} shows a faster learning curve than \textit{Hed-HC}. Also, it is observed that sending soft information to FC helps to further reduce the collision with PUs and other SUs which is of our great interest by keeping missed opportunity at the nearly same level. Both \textit{Hed-SC} and \textit{Perc-SC} show improvement over \textit{Hed-HC}. Since this is an online learning approach, a transient behavior is observed during the initial steps as seen in Fig. \ref{fig:50su_pktcoll_MSC} and \ref{fig:50su_int_MSC}. The PU interference observed during the transmission depends also on the density of SUs in the CRN (See Fig. \ref{fig:50su_int_MSC} and \ref{fig:10su_int_BSC}). This is because, with more devices, the fraction of devices satisfying $P_d>0.95$ is higher and hence the FC can detect the state of the channel more accurately. Also note that the PU interference of \textit{AND}\cite{1542650} and \textit{CV}\cite{4446521} is high because in an IoT scenario with wide geo-deployment as considered, there will always be some SU which will be in deep-fade/shadow and will not be able to detect the presence of PU accurately.
\revAdd{We can see that the proposed Hedge method is able to provide better performance than the computationally complex perceptron method.}

We can see from Fig. \ref{fig:Comp_with_traditional} that proposed methods perform very close to the offline pre-trained deep network in all the metrics. We report the average behavior of the DL based method. Note that this performance of the proposed methods is obtained by learning online with approximate ground truth whereas the DL methods are pre-trained with actual ground truths. This improved performance of the proposed approach is because of including domain knowledge while computing the thresholds as given in (\ref{eqn:gamma}) and (\ref{eqn:gamma_p}). However, DL methods are trained without any domain knowledge. Training DL methods with domain knowledge are not well established and is an interesting future work. A recent development in Online deep learning \cite{sahoo2017online} enables the training of models in streaming data. However, training them with approximate ground truth is challenging and could be an interesting future direction.

\revAdd{In Fig. \ref{fig5} we provide the ROC curve in medium signal condition for the Hedge hard combining (``Hed-HC") and Hedge soft combining (``Hed-SC") algorithms. The plotted $P_d$ and $P_{fa}$ are empirically calculated at fusion center and we can see that ROC for ``Hed-SC" lies above ``Hed-HC" as expected.}

In Fig. \ref{fig:switch} the results for comparing unmodified Hedge and Perceptron algorithm against the algorithms with BH procedure is presented. \textit{Hed-SC} and \textit{Perc-SC} with BH procedure are called \textit{"HSC-BH"} and \textit{"PSC-BH"}. The switch method between \textit{Hed-SC} and \textit{HSC-BH} is called \textit{"HSC-sw"} and same switch method for perceptron is called \textit{"PSC-sw"}. In poor signal condition \textit{Hed-SC} and \textit{Perc-SC} perform better as shown in Fig. \ref{fig:switch} but in GSC (shown in appendix) and MSC, \textit{HSC-BH} and \textit{PSC-BH} algorithms reduces the missed slots for transmission considerably with negligible increase in SU and PU collision. As proposed in section \ref{switch_BH_traditional}, we show here that \textit{HSC-sw} and \textit{PSC-sw} perform well in all signal conditions. It minimizes the missed slots for transmission keeping collision at each SU under a value $\tau = 0.02$. 

\setcounter{figure}{3}
\begin{figure}[!ht]
    \centering
    \begin{subfigure}{.25\textwidth}
        \includegraphics[width=\linewidth]{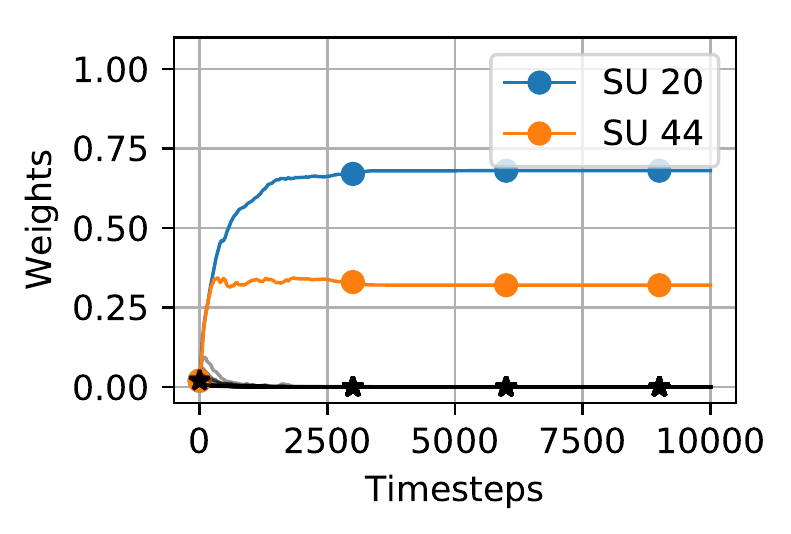}
        \caption{Hedge hard combining}
    \end{subfigure}%
    \begin{subfigure}{.25\textwidth}
        \includegraphics[width=\linewidth]{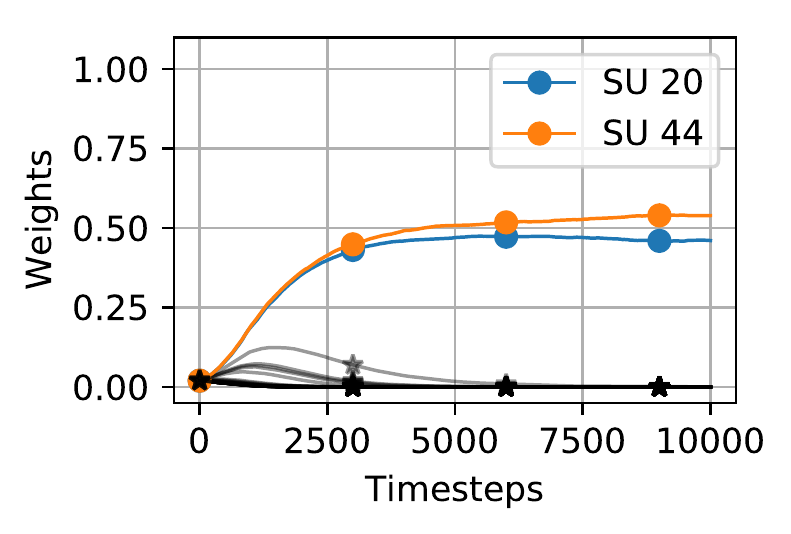}
        \caption{Hedge soft combining}
    \end{subfigure}
    \caption{Weight evolution in MSC for stationary channel condition}
    \label{fig3}
    \end{figure}

\revAdd{As the learning progresses, the FC can observe which SUs are successful at reporting the correct channel state for each PUs. However, toward later stages, these SUs will be given higher weight. As transmission progresses, the weights assigned to SUs will converge under assumed conditions of stationarity. By simulation, we show that the hedge weights for any particular PU converge. We consider a scenario with $50$ SUs where two SUs (SU$20$ and SU$44$) can almost perfectly sense the occupancy of PU and almost all other SUs are blind to the PU. The normalized weight evolution of each SUs in the system is given in Fig. \ref{fig3}. We can observe that SU$20$ and SU$44$ are having non-zero weights while all others are close to zero. This implies that for the final decision about the channel state of a PU, the decisions from SU$20$ and SU$44$ will be given more importance at the fusion center.}

\ifCLASSOPTIONonecolumn
\begin{figure}[!ht]
\centering
    \begin{subfigure}{.33\textwidth}
        \includegraphics[width=1.\linewidth]{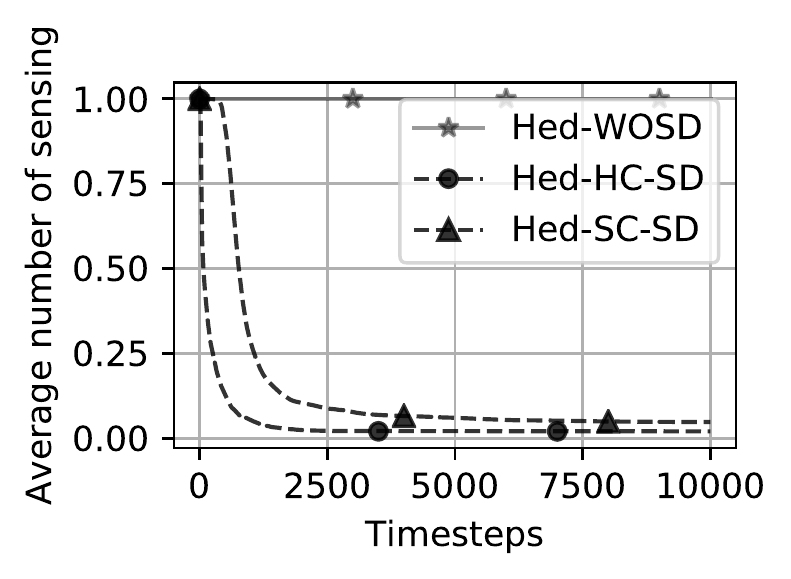}
        \caption{MSC}
        \label{fig:NS10SU}
    \end{subfigure}%
    \begin{subfigure}{.33\textwidth}
        \includegraphics[width=1.\linewidth]{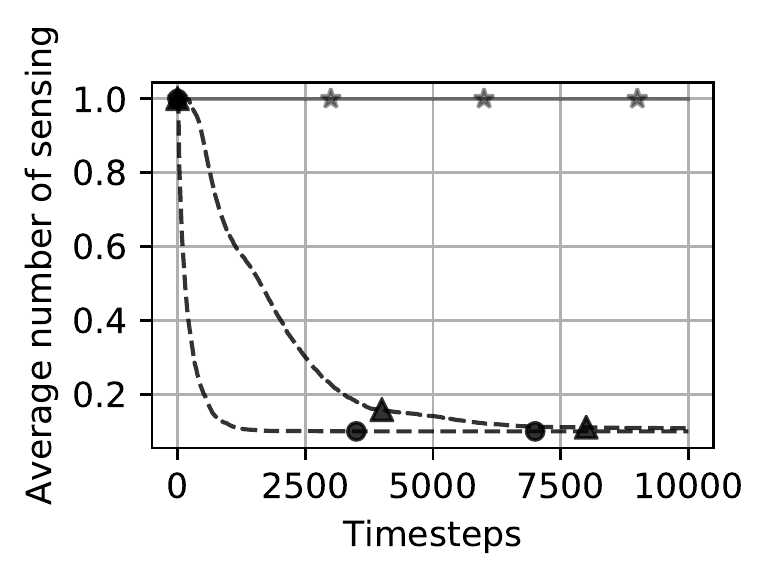}
        \caption{BSC}
        \label{fig:NS50SU}
    \end{subfigure}
    \caption{Average number of sensing per SU per timestep}
    \label{fig:comp31}
\end{figure}
\else
\begin{figure}[!ht]
\centering
    \begin{subfigure}{.25\textwidth}
        \includegraphics[width=1.\linewidth]{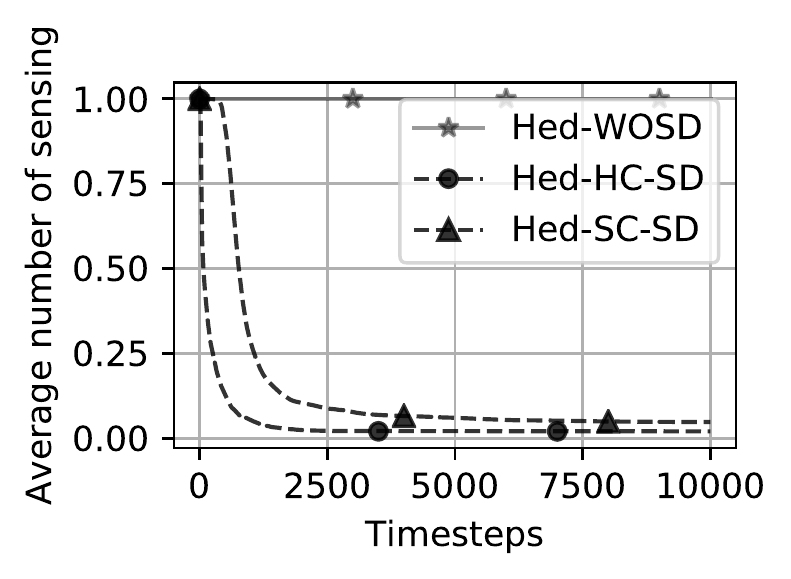}
        \caption{MSC}
        \label{fig:NS10SU}
    \end{subfigure}%
    \begin{subfigure}{.25\textwidth}
        \includegraphics[width=1.\linewidth]{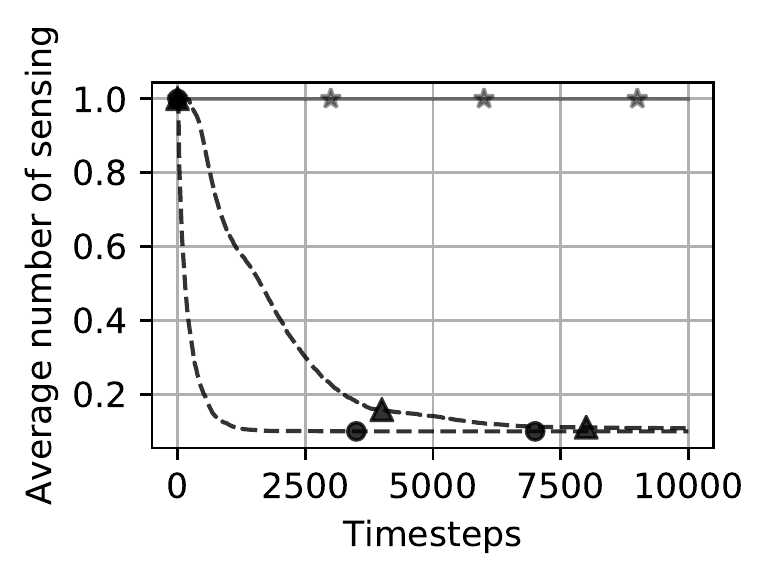}
        \caption{BSC}
        \label{fig:NS50SU}
    \end{subfigure}
    \caption{Average number of sensing per SU per timestep}
    \label{fig:comp31}
\end{figure}
\fi

\setcounter{figure}{7}
\begin{figure*}[!ht]
    \centering
    \begin{subfigure}{.33\textwidth}
        \includegraphics[width=\linewidth]{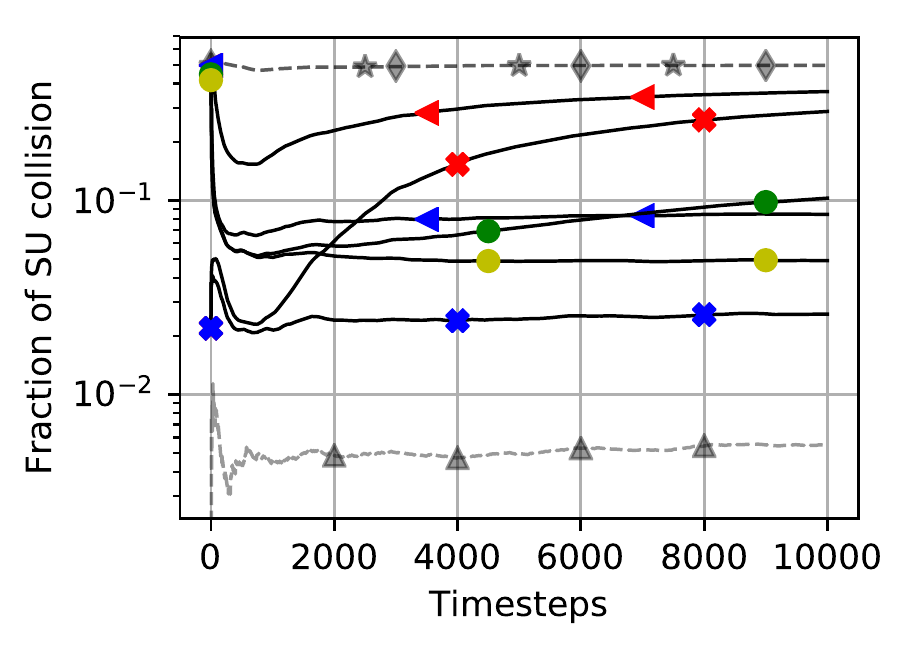}
        \caption{Fraction of SU packet collision}
    \end{subfigure}%
    \begin{subfigure}{.33\textwidth}
        \includegraphics[width=\linewidth]{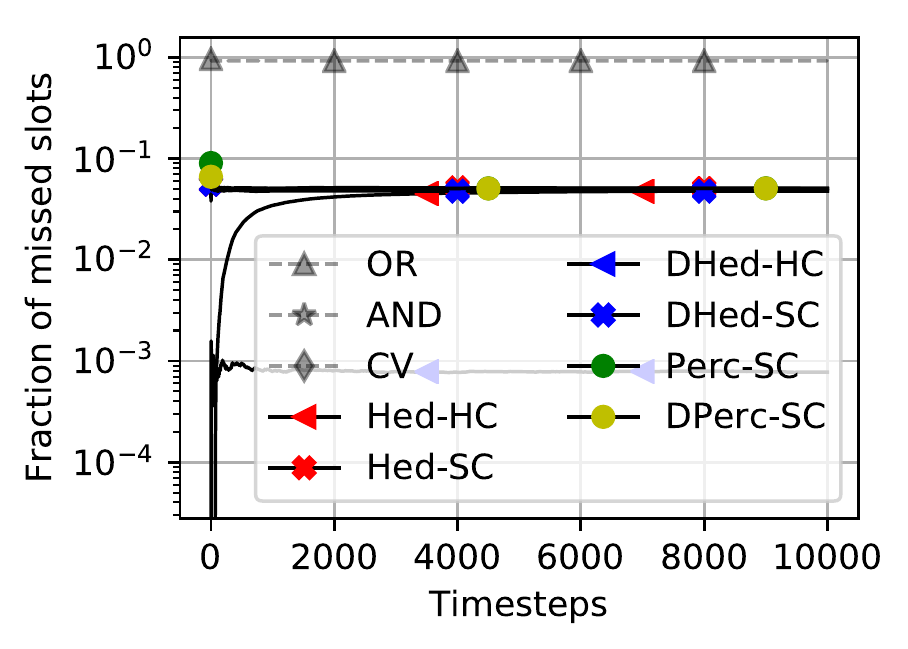}
        \caption{Fraction of missed idle slots}
    \end{subfigure}%
    \begin{subfigure}{.33\textwidth}
        \includegraphics[width=\linewidth]{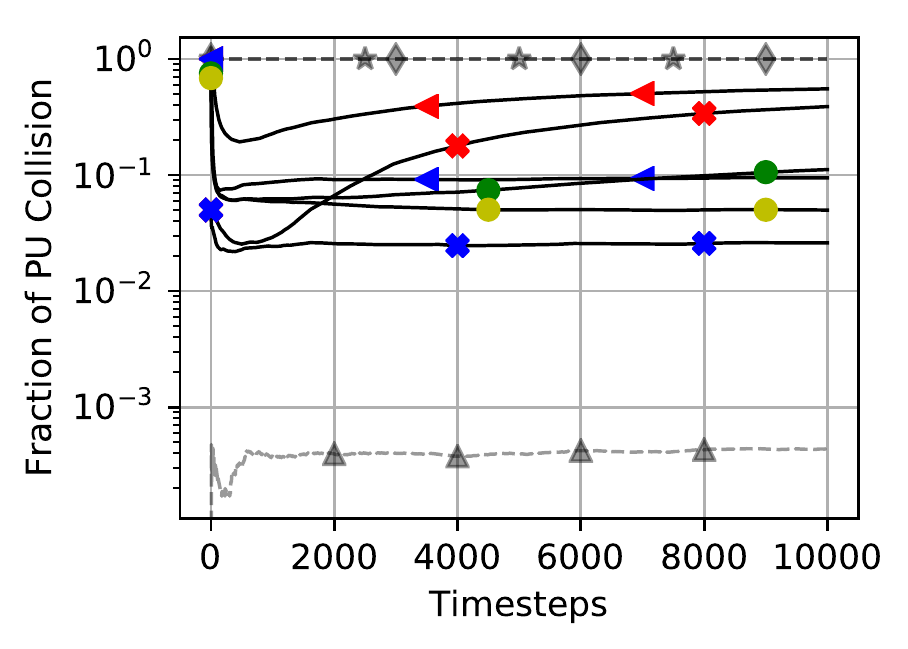}
        \caption{Observed interference at PU}
    \end{subfigure}
    \caption{Both PUs and SUs are mobile in Medium Signal Condition}
    \label{fig12}
\end{figure*}

\begin{figure*}[!ht]
    \centering
    \begin{subfigure}{.33\textwidth}
        \includegraphics[width=\linewidth]{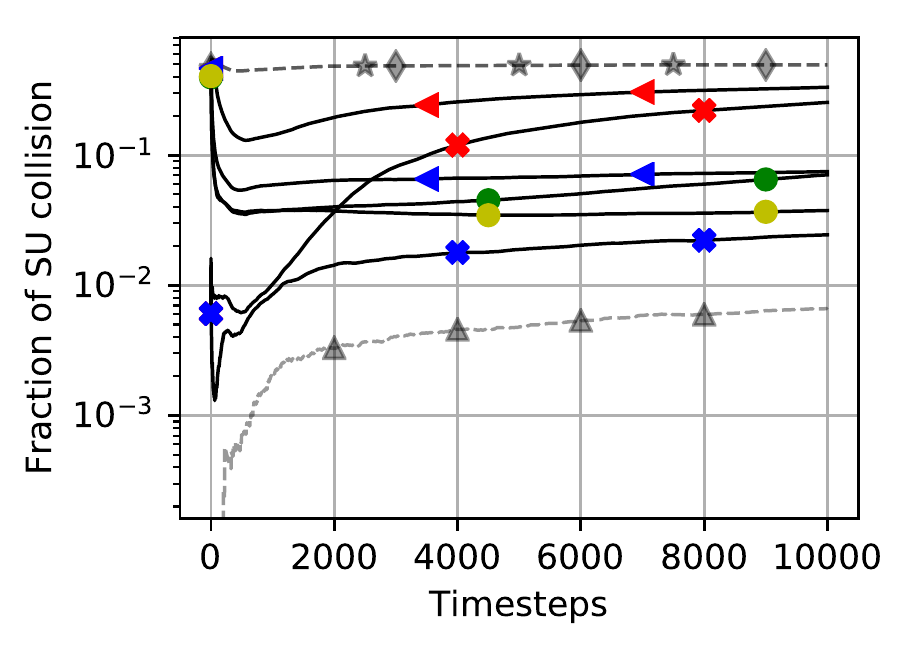}
        \caption{Fraction of SU packet collision}
    \end{subfigure}%
    \begin{subfigure}{.33\textwidth}
        \includegraphics[width=\linewidth]{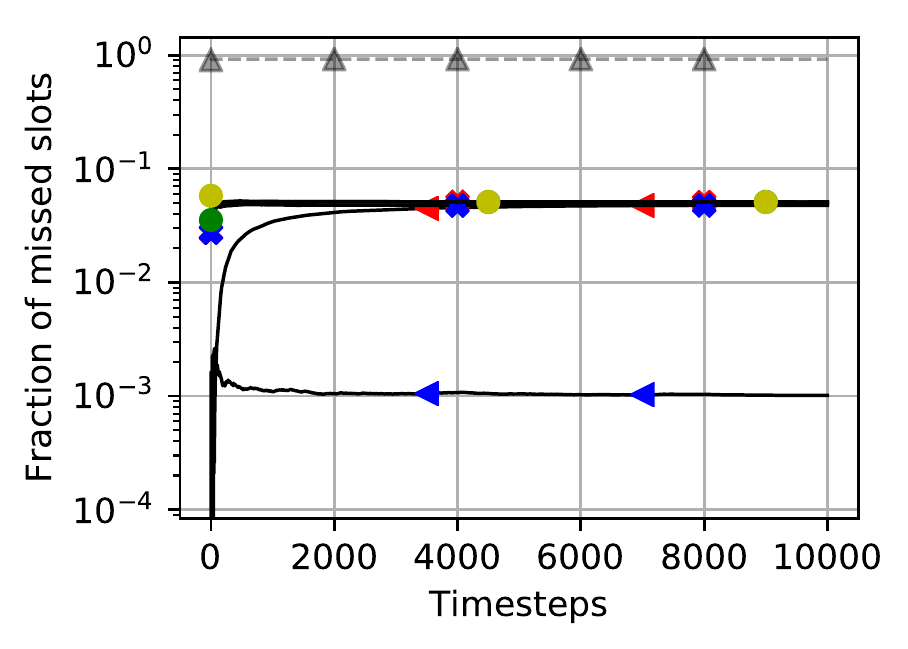}
        \caption{Fraction of missed idle slots}
    \end{subfigure}%
    \begin{subfigure}{.33\textwidth}
        \includegraphics[width=\linewidth]{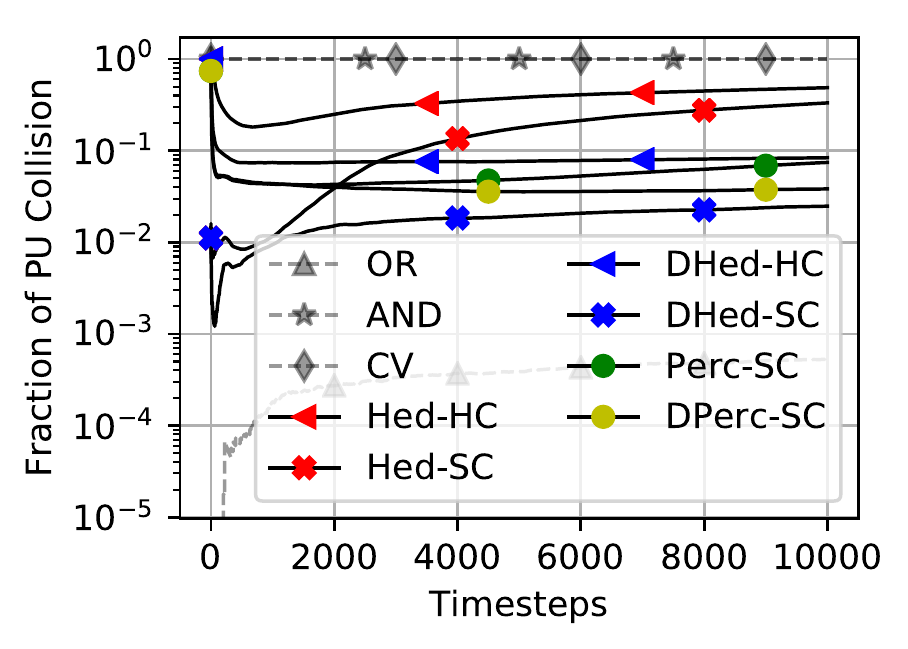}
        \caption{Observed interference at PU}
    \end{subfigure}
    \caption{Mobile PUs with stationary SUs in Medium Signal Condition}
    \label{fig11}
\end{figure*}

From Fig. \ref{fig:NS10SU} - \ref{fig:NS50SU} we observe that the number of sensing can be decreased by selectively deactivating the poor-performing SUs in Hedge algorithm (SUs with low normalized weights). This is achieved with almost no degradation in collision and missed slots for transmission as shown in Fig. \ref{fig:50SU_result_SD}. The label \textit{"Hed-WOSD"} represents the Hedge algorithm without deactivating any SU and, \textit{"Hed-HC-SD"} and \textit{"Hed-SC-SD"} corresponds to the Hedge algorithms (Hard Combining and Soft Combining) with selectively deactivating SUs as discussed in Sec. \ref{sec:energy_saving}. This is because the SUs with low weights do not contribute much to the final decision made by FC and hence the loss stays almost the same.

To observe the effect of selectively deactivating detectors from sensing poor channels on energy savings, we consider an energy limited scenario with all devices have a budget of $10000$ units of energy in a $10$ PU.

For simplicity, we assume one sensing operation requires $1$ unit of energy i.e if an SU senses all channels it spends 10 units in a time step. 

\setcounter{figure}{6}
\ifCLASSOPTIONonecolumn
\begin{figure}[!ht]
\centering
    \begin{subfigure}{.33\textwidth}
        \includegraphics[width=1.\linewidth]{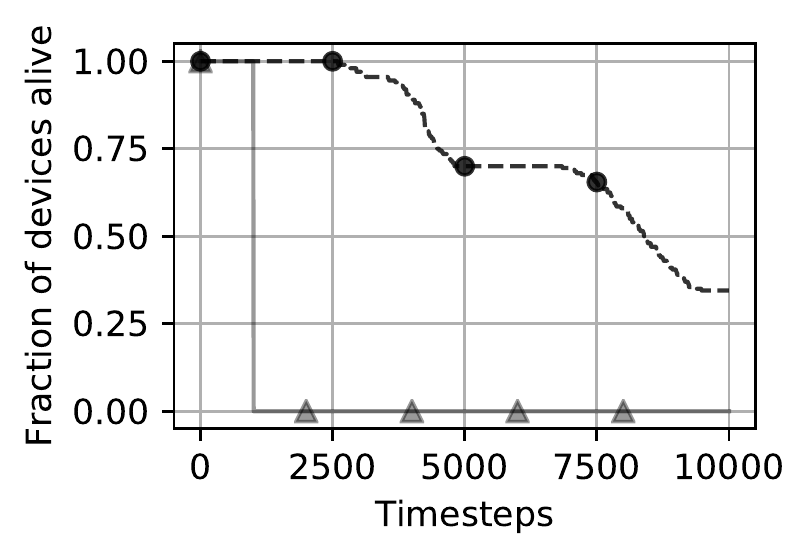}
        \caption{BSC}
        \label{fig:dev_left10SU}
    \end{subfigure}%
    \begin{subfigure}{.33\textwidth}
        \includegraphics[width=1.\linewidth]{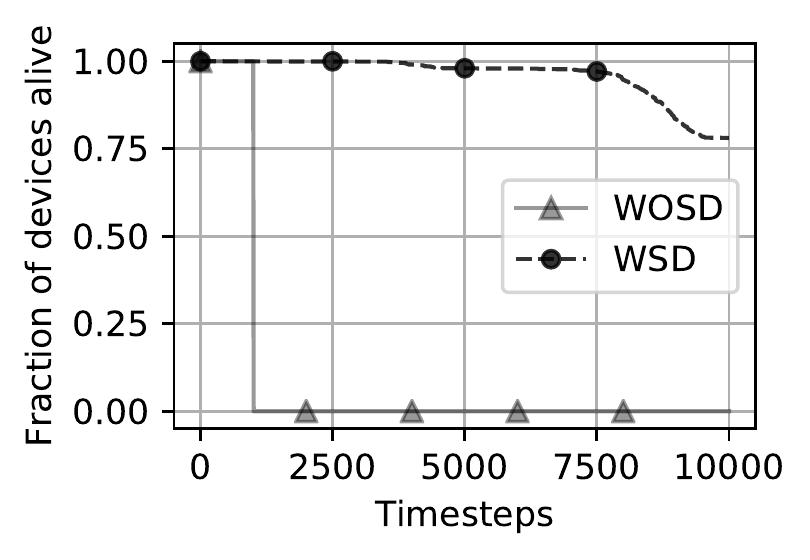}
        \caption{MSC}
        \label{fig:dev_left50SU}
    \end{subfigure}
    \caption{Comparison of fraction of devices left with energy}
    \label{fig:comp4}
\end{figure}
\else
\begin{figure}[!ht]
\centering
    \begin{subfigure}{.25\textwidth}
        \includegraphics[width=1.\linewidth]{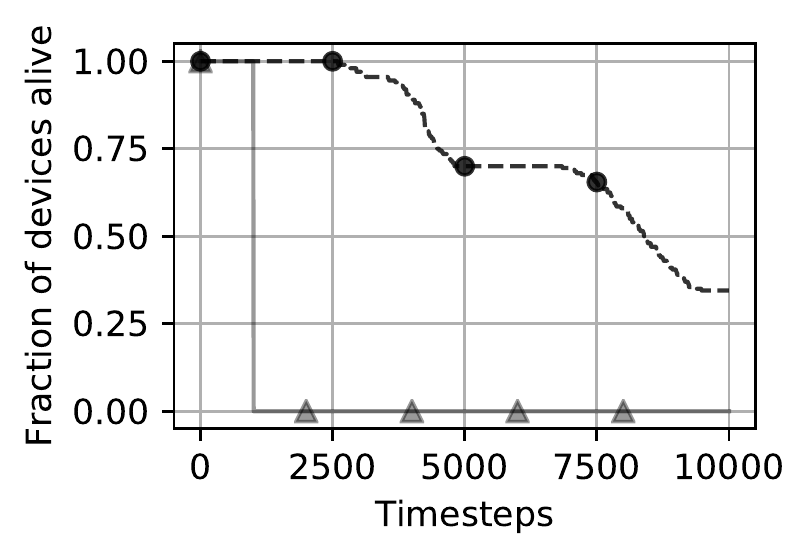}
        \caption{BSC}
        \label{fig:dev_left10SU}
    \end{subfigure}%
    \begin{subfigure}{.25\textwidth}
        \includegraphics[width=1.\linewidth]{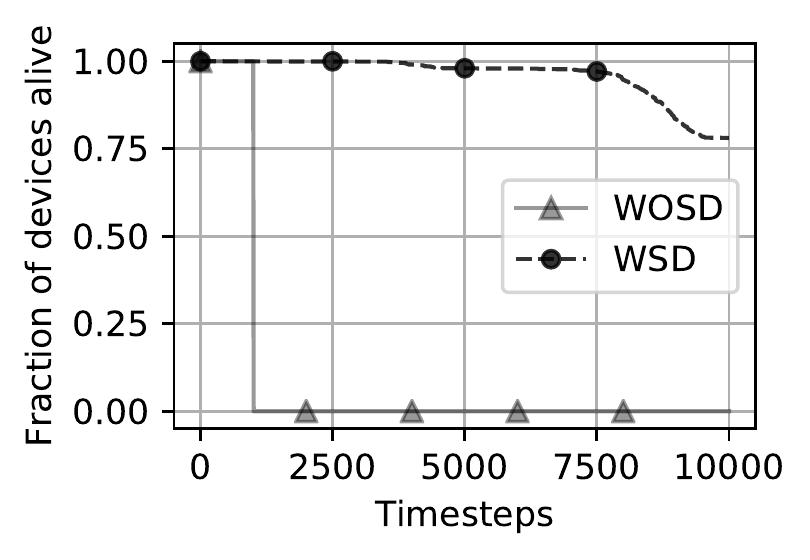}
        \caption{MSC}
        \label{fig:dev_left50SU}
    \end{subfigure}
    \caption{Comparison of fraction of devices left with energy}
    \label{fig:comp4}
\end{figure}
\fi

We do not account for the energy expenditure for transmission as we aim to analyze the effect of energy savings by eliminating sensing operations for some SUs and the corresponding advantage in extending the field life of devices. We use $\mu = \frac{1}{2S}$ as the cut-off threshold on normalized weights for deactivating the SUs. We provide the fraction of devices left alive after each time step in Fig. \ref{fig:comp4}. In figure, \textit{"WOSD"} refers fraction of devices left without selectively deactivating any SUs and "\textit{WSD}" refers same with deactivating poor-performing SUs. With $10000$ units of energy as a budget to each SU, traditional approaches which do not deactivate any SUs can survive only for $1000$ time steps as it spends $10$ units of energy to sense the available $10$ channels at each step.  
However, the proposed online learning method with the deactivation of detectors based on normalized weight can extend the lifetime of the devices without compromising on sensing performance as seen from Fig. \ref{fig:50SU_result_SD}. Also from Fig. \ref{fig:comp4} we can note that with the increasing number of SUs, the proposed approach can conserve energy for more number of devices for a longer duration. 

Note that this method cannot be applied to DL methods as the DL models require the same dimension of input always. Switching off will change the dimension of the input of the prediction models rendering DL methods unusable and this can happen often in dynamic environments.

\setcounter{figure}{9}
\begin{figure*}[ht]
    \centering
    \begin{subfigure}{.33\textwidth}
        \includegraphics[width=\linewidth]{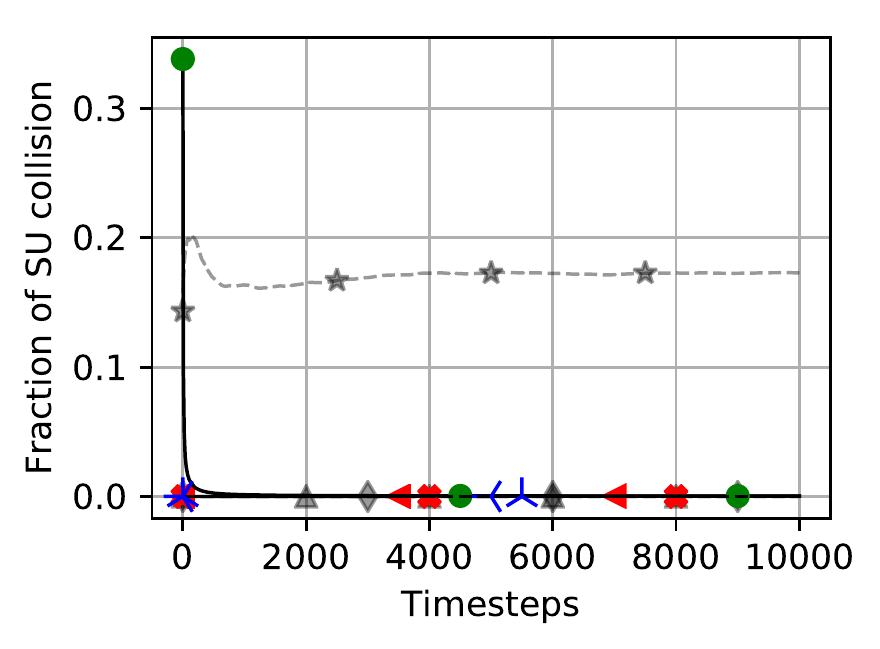}
        \caption{Fraction of SU packet collision in GSC}
        \label{fig:10su_pktcoll_GSC}
    \end{subfigure}%
    \begin{subfigure}{.33\textwidth}
        \includegraphics[width=\linewidth]{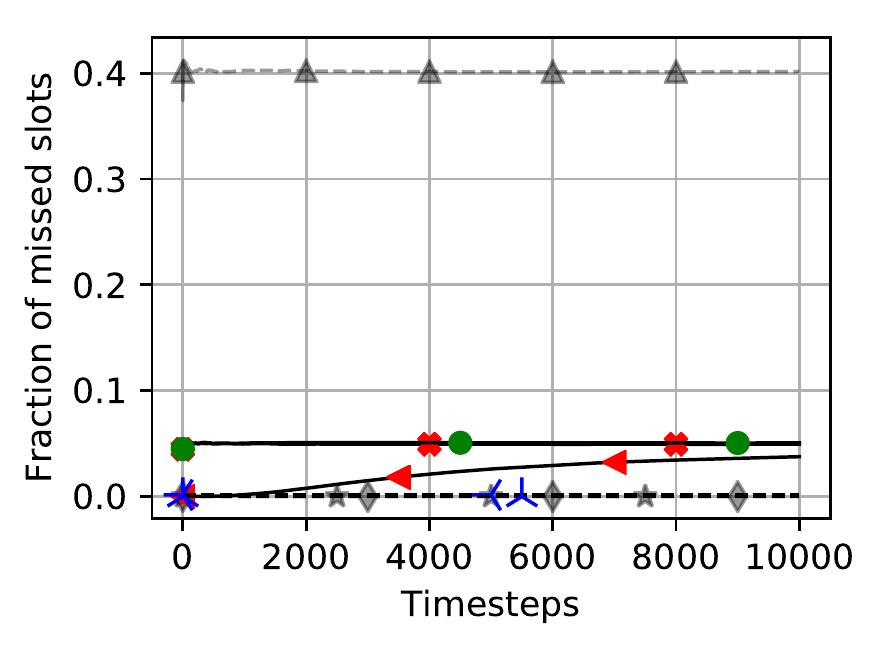}
        \caption{Fraction of missed idle slots in GSC}
        \label{fig:10su_missed_GSC}
    \end{subfigure}%
    \begin{subfigure}{.33\textwidth}
        \includegraphics[width=\linewidth]{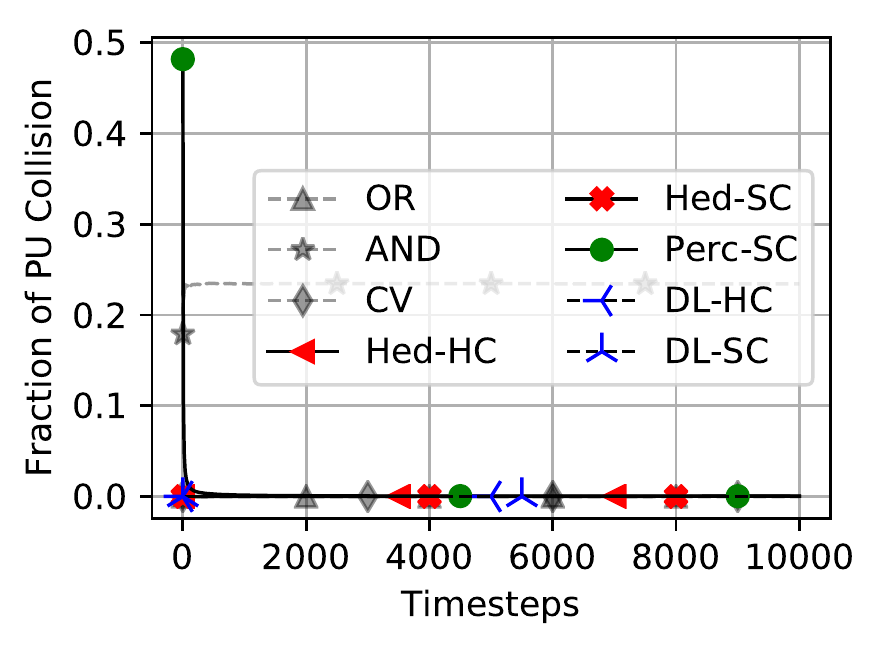}
        \caption{Observed interference at PU in GSC}
        \label{fig:10su_int_GSC}
    \end{subfigure}
    \caption{Three metrics are compared for proposed Hedge-HC, Hedge-SC and Perceptron-SC with traditional OR, AND and CV for GSC}
    \label{fig:App_Comp_with_traditional}
\end{figure*}

\begin{figure*}[ht]
    \centering
    \begin{subfigure}{.33\textwidth}
        \includegraphics[width=\linewidth]{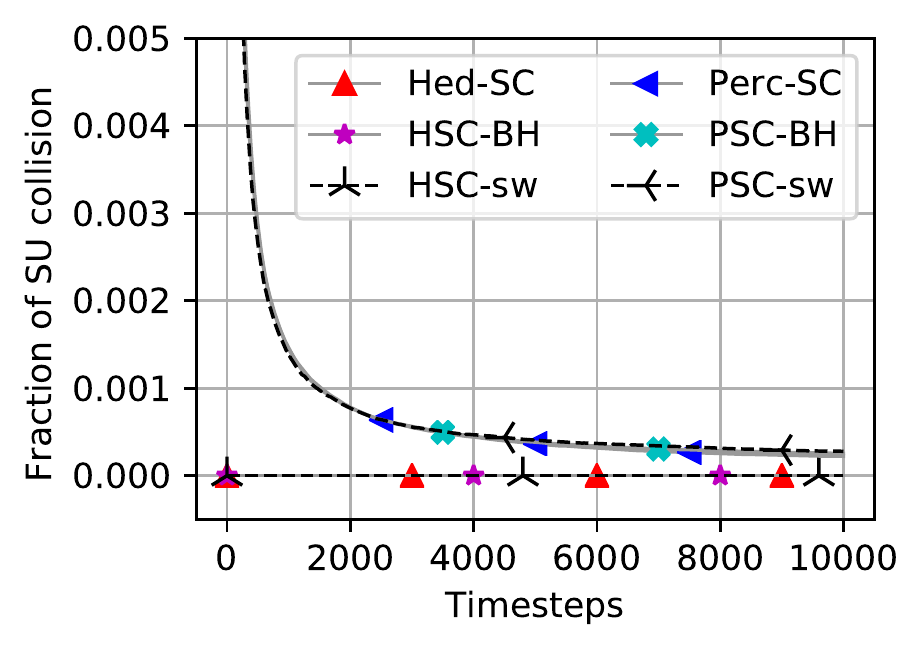}
        \caption{Fraction of SU packet collision for GSC}
        \label{fig:10su_pktcoll}
    \end{subfigure}%
    \begin{subfigure}{.33\textwidth}
        \includegraphics[width=\linewidth]{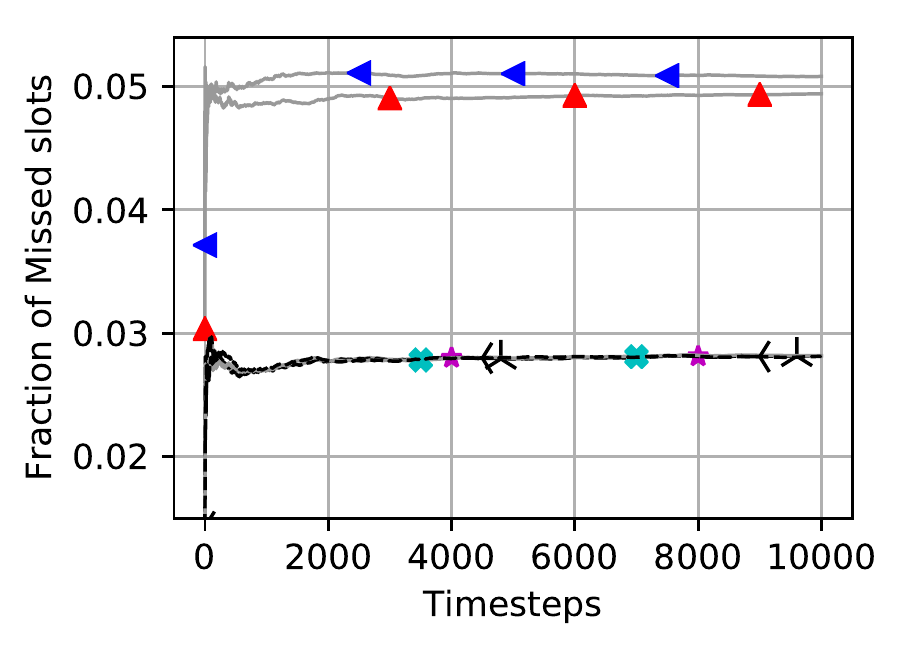}
        \caption{Fraction of missed idle slots for GSC}
        \label{fig:10su_missed}
    \end{subfigure}%
    \begin{subfigure}{.33\textwidth}
        \includegraphics[width=\linewidth]{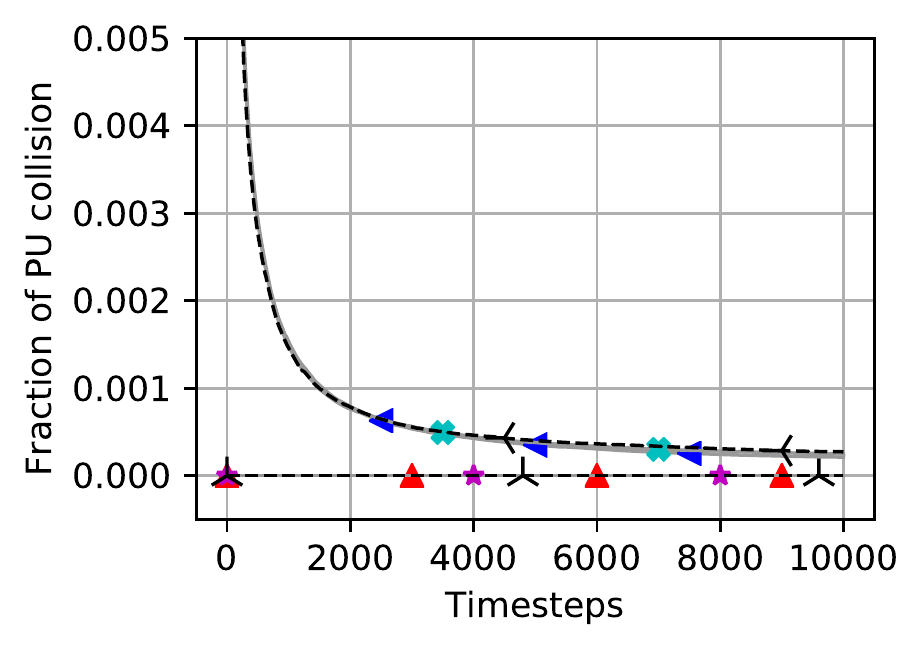}
        \caption{Observed interference at PU for GSC}
    \end{subfigure}
    \caption{Comparison of switch strategy in good signal conditions}
    \label{fig:App_switch}
\end{figure*}

\revAdd{The performance of dHedge and dPerceptron for the two cases of non-stationarity i.e. (i) both PU and SUs are mobile and (ii) only PUs are mobile are shown in Fig. \ref{fig12} and Fig. \ref{fig11} for medium signal condition. However, both dHedge and dPerceptron are equally applicable in the cases of good and bad signal conditions as well. In the simulation, we have considered MSC with 50 SUs where PU and/or SUs move with a velocity of $5$m/s.
We used $\gamma=0.80, \beta=0.05$ for dHedge hard combining, $\gamma=0.60, \beta=0.50$ for dHedge soft combining and $\gamma=0.99, \rho=0.40$ for dPerceptron and in medium signal condition. For both of the cases of non-stationarity, the PU and SU collisions and missed slots show increasing trends over time using Hedge and Perceptron as given by ``Hed-HC", ``Hed-SC" and ``Perc-SC" in Fig. \ref{fig12} and Fig. \ref{fig11}. But using dHedge and dPerceptron, labeled ``DHed-HC", ``DHed-SC" and ``DPerc-SC", shows improved performance in all the three metrics considered.}

\section{Conclusion}
This paper presented a learning-based approach to tackle the problem of CSS in IoT networks. In summary, we presented an online learning framework for collaborative spectrum sensing which (i) learns to combine the information from different sensing devices according to their quality of reporting, (ii) can save energy of the devices and thus extend the lifetime of the network by selectively switching off the sensors, (iii) can be easily scaled to networks experiencing wide variety of signal conditions and large number of devices and (iv) can be easily modified to handle situations where devices drop-out of the network randomly. In an IoT scenario, typically  (i) SUs with varying sensing capabilities are present, (ii) energy efficiency is crucial, (iii) scalability is necessary and (iv) the devices can join and leave the network in an ad-hoc manner. 
\revAdd{Though the proposed method only deals with improving the channel sensing capability, it can be combined with other existing scheduling mechanisms and may be able to provide improved transmission capabilities. Also, the information from the proposed algorithms (e.g. weights) can be exploited to design specialized schedulers which could holistically improve idle spectrum utilization. This could be an interesting future work.}

    \appendix

In good signal condition (GSC) online and DL methods are compared concerning three metrics in Fig. \ref{fig:App_Comp_with_traditional}. We can observe that both HC and SC are performing equally good and very competitive to offline trained deep network. The effect of switching between traditional online learning and online learning with BH in GSC is shown in Fig. \ref{fig:App_switch}. Note that in GSC, \textit{Switch-BH} follows the online method with the BH and can considerably reduce the fraction of missed transmission opportunity.

	\bibliographystyle{IEEEtran}
 	\bibliography{99_library.bib}

\begin{thebibliography}{10}
\providecommand{\url}[1]{#1}
\csname url@samestyle\endcsname
\providecommand{\newblock}{\relax}
\providecommand{\bibinfo}[2]{#2}
\providecommand{\BIBentrySTDinterwordspacing}{\spaceskip=0pt\relax}
\providecommand{\BIBentryALTinterwordstretchfactor}{4}
\providecommand{\BIBentryALTinterwordspacing}{\spaceskip=\fontdimen2\font plus
\BIBentryALTinterwordstretchfactor\fontdimen3\font minus
  \fontdimen4\font\relax}
\providecommand{\BIBforeignlanguage}[2]{{%
\expandafter\ifx\csname l@#1\endcsname\relax
\typeout{** WARNING: IEEEtran.bst: No hyphenation pattern has been}%
\typeout{** loaded for the language `#1'. Using the pattern for}%
\typeout{** the default language instead.}%
\else
\language=\csname l@#1\endcsname
\fi
#2}}
\providecommand{\BIBdecl}{\relax}
\BIBdecl

\bibitem{GUBBI20131645}
J.~Gubbi, R.~Buyya, S.~Marusic, and M.~Palaniswami, ``Internet of things (iot):
  A vision, architectural elements, and future directions,'' \emph{Future
  Generation Computer Systems}, vol.~29, no.~7, pp. 1645 -- 1660, 2013.

\bibitem{ATZORI20102787}
L.~Atzori, A.~Iera, and G.~Morabito, ``The internet of things: A survey,''
  \emph{Computer Networks}, vol.~54, no.~15, pp. 2787 -- 2805, 2010.

\bibitem{6710070}
L.~{Atzori}, A.~{Iera}, and G.~{Morabito}, ``From "smart objects" to "social
  objects": The next evolutionary step of the internet of things,'' \emph{IEEE
  Communications Magazine}, vol.~52, no.~1, pp. 97--105, January 2014.

\bibitem{788210}
J.~{Mitola} and G.~Q. {Maguire}, ``Cognitive radio: making software radios more
  personal,'' \emph{IEEE Personal Communications}, vol.~6, no.~4, pp. 13--18,
  Aug 1999.

\bibitem{7006643}
A.~{Aijaz} and A.~H. {Aghvami}, ``Cognitive machine-to-machine communications
  for internet-of-things: A protocol stack perspective,'' \emph{IEEE Internet
  of Things Journal}, vol.~2, no.~2, pp. 103--112, April 2015.

\bibitem{8270374}
V.~{Raj}, I.~{Dias}, T.~{Tholeti}, and S.~{Kalyani}, ``Spectrum access in
  cognitive radio using a two-stage reinforcement learning approach,''
  \emph{IEEE Journal of Selected Topics in Signal Processing}, vol.~12, no.~1,
  pp. 20--34, Feb 2018.

\bibitem{DBLP:journals/corr/abs-1804-11135}
T.~Tholeti, V.~Raj, and S.~Kalyani, ``A non-parametric multi-stage learning
  framework for cognitive spectrum access in iot networks,'' \emph{CoRR}, vol.
  abs/1804.11135, 2018.

\bibitem{bayhan2013scheduling}
S.~Bayhan and F.~Alagoz, ``Scheduling in centralized cognitive radio networks
  for energy efficiency,'' \emph{IEEE Transactions on Vehicular Technology},
  vol.~62, no.~2, pp. 582--595, 2013.

\bibitem{zhang2018artificial}
K.~{Zhang}, S.~{Leng}, X.~{Peng}, L.~{Pan}, S.~{Maharjan}, and Y.~{Zhang},
  ``Artificial intelligence inspired transmission scheduling in cognitive
  vehicular communications and networks,'' \emph{IEEE Internet of Things
  Journal}, vol.~6, no.~2, pp. 1987--1997, April 2019.

\bibitem{4796930}
T.~{Yucek} and H.~{Arslan}, ``A survey of spectrum sensing algorithms for
  cognitive radio applications,'' \emph{IEEE Communications Surveys Tutorials},
  vol.~11, no.~1, pp. 116--130, First 2009.

\bibitem{1399240}
D.~{Cabric}, S.~M. {Mishra}, and R.~W. {Brodersen}, ``Implementation issues in
  spectrum sensing for cognitive radios,'' in \emph{Conference Record of the
  Thirty-Eighth Asilomar Conference on Signals, Systems and Computers, 2004.},
  vol.~1, Nov 2004, pp. 772--776 Vol.1.

\bibitem{1542627}
A.~{Ghasemi} and E.~S. {Sousa}, ``Collaborative spectrum sensing for
  opportunistic access in fading environments,'' in \emph{First IEEE
  International Symposium on New Frontiers in Dynamic Spectrum Access Networks,
  2005. DySPAN 2005.}, Nov 2005, pp. 131--136.

\bibitem{1542650}
E.~{Visotsky}, S.~{Kuffner}, and R.~{Peterson}, ``On collaborative detection of
  tv transmissions in support of dynamic spectrum sharing,'' in \emph{First
  IEEE International Symposium on New Frontiers in Dynamic Spectrum Access
  Networks, 2005. DySPAN 2005.}, Nov 2005, pp. 338--345.

\bibitem{4446521}
C.~Lee and W.~Wolf, ``Energy efficient techniques for cooperative spectrum
  sensing in cognitive radios,'' in \emph{2008 5th IEEE Consumer Communications
  and Networking Conference}, Jan 2008, pp. 968--972.

\bibitem{4024390}
S.~M. {Mishra}, A.~{Sahai}, and R.~W. {Brodersen}, ``Cooperative sensing among
  cognitive radios,'' in \emph{2006 IEEE International Conference on
  Communications}, vol.~4, June 2006, pp. 1658--1663.

\bibitem{7161310}
G.~{Chandrasekaran} and S.~{Kalyani}, ``Performance analysis of cooperative
  spectrum sensing over$\kappa{-}\mu$shadowed fading,'' \emph{IEEE Wireless
  Communications Letters}, vol.~4, no.~5, pp. 553--556, Oct 2015.

\bibitem{4686831}
J.~{Ma}, G.~{Zhao}, and Y.~{Li}, ``Soft combination and detection for
  cooperative spectrum sensing in cognitive radio networks,'' \emph{IEEE
  Transactions on Wireless Communications}, vol.~7, no.~11, pp. 4502--4507,
  November 2008.

\bibitem{6682626}
W.~{Choi}, M.~{Song}, J.~{Ahn}, and G.~{Im}, ``Soft combining for cooperative
  spectrum sensing over fast-fading channels,'' \emph{IEEE Communications
  Letters}, vol.~18, no.~2, pp. 193--196, February 2014.

\bibitem{benjamini1995controlling}
Y.~Benjamini and Y.~Hochberg, ``Controlling the false discovery rate: a
  practical and powerful approach to multiple testing,'' \emph{Journal of the
  Royal statistical society: series B (Methodological)}, vol.~57, no.~1, pp.
  289--300, 1995.

\bibitem{4446520}
G.~{Atia}, S.~{Aeron}, E.~{Ermis}, and V.~{Saligrama}, ``On throughput
  maximization and interference avoidance in cognitive radios,'' in \emph{2008
  5th IEEE Consumer Communications and Networking Conference}, Jan 2008, pp.
  963--967.

\bibitem{6635250}
K.~M. {Thilina}, K.~W. {Choi}, N.~{Saquib}, and E.~{Hossain}, ``Machine
  learning techniques for cooperative spectrum sensing in cognitive radio
  networks,'' \emph{IEEE Journal on Selected Areas in Communications}, vol.~31,
  no.~11, pp. 2209--2221, November 2013.

\bibitem{8302117}
D.~{Han}, G.~C. {Sobabe}, C.~{Zhang}, X.~{Bai}, Z.~{Wang}, S.~{Liu}, and
  B.~{Guo}, ``Spectrum sensing for cognitive radio based on convolution neural
  network,'' in \emph{2017 10th International Congress on Image and Signal
  Processing, BioMedical Engineering and Informatics (CISP-BMEI)}, Oct 2017,
  pp. 1--6.

\bibitem{8604101}
W.~{Lee}, M.~{Kim}, and D.~{Cho}, ``Deep cooperative sensing: Cooperative
  spectrum sensing based on convolutional neural networks,'' \emph{IEEE
  Transactions on Vehicular Technology}, vol.~68, no.~3, pp. 3005--3009, March
  2019.

\bibitem{lees2019deep}
W.~M. Lees, A.~Wunderlich, P.~Jeavons, P.~D. Hale, and M.~R. Souryal, ``Deep
  learning classification of 3.5 ghz band spectrograms with applications to
  spectrum sensing,'' \emph{IEEE Transactions on Cognitive Communications and
  Networking}, 2019.

\bibitem{arjoune2019comprehensive}
Y.~Arjoune and N.~Kaabouch, ``A comprehensive survey on spectrum sensing in
  cognitive radio networks: Recent advances, new challenges, and future
  research directions,'' \emph{Sensors}, vol.~19, no.~1, p. 126, 2019.

\bibitem{zhang2010optimal}
T.~Zhang, Y.~Wu, K.~Lang, and D.~H. Tsang, ``Optimal scheduling of cooperative
  spectrum sensing in cognitive radio networks,'' \emph{IEEE Systems Journal},
  vol.~4, no.~4, pp. 535--549, 2010.

\bibitem{zhao2007decentralized}
Q.~Zhao, L.~Tong, A.~Swami, and Y.~Chen, ``Decentralized cognitive mac for
  opportunistic spectrum access in ad hoc networks: A pomdp framework,''
  CALIFORNIA UNIV DAVIS DEPT OF ELECTRICAL AND COMPUTER ENGINEERING, Tech.
  Rep., 2007.

\bibitem{raschella2013use}
A.~Raschell{\`a}, J.~P{\'e}rez-Romero, O.~Sallent, and A.~Umbert, ``On the use
  of pomdp for spectrum selection in cognitive radio networks,'' in \emph{8th
  International Conference on Cognitive Radio Oriented Wireless
  Networks}.\hskip 1em plus 0.5em minus 0.4em\relax IEEE, 2013, pp. 19--24.

\bibitem{lee2011enhanced}
W.~Lee and D.-H. Cho, ``Enhanced spectrum sensing scheme in cognitive radio
  systems with mimo antennae,'' \emph{IEEE transactions on vehicular
  technology}, vol.~60, no.~3, pp. 1072--1085, 2011.

\bibitem{freund1997decision}
Y.~Freund and R.~E. Schapire, ``A decision-theoretic generalization of on-line
  learning and an application to boosting,'' \emph{Journal of computer and
  system sciences}, vol.~55, no.~1, pp. 119--139, 1997.

\bibitem{Freund1999}
\BIBentryALTinterwordspacing
------, ``Large margin classification using the perceptron algorithm,''
  \emph{Machine Learning}, vol.~37, no.~3, pp. 277--296, Dec 1999. [Online].
  Available: \url{https://doi.org/10.1023/A:1007662407062}
\BIBentrySTDinterwordspacing

\bibitem{goldsmith2009breaking}
A.~J. Goldsmith, S.~A. Jafar, I.~Maric, and S.~Srinivasa, ``Breaking spectrum
  gridlock with cognitive radios: An information theoretic perspective.''
  \emph{Proceedings of the IEEE}, vol.~97, no.~5, pp. 894--914, 2009.

\bibitem{singh2015cooperative}
J.~S.~P. Singh, R.~Singh, M.~K. Rai, J.~Singh, and A.~Kang, ``Cooperative
  sensing for cognitive radio: A powerful access method for shadowing
  environment,'' \emph{Wireless Personal Communications}, vol.~80, no.~4, pp.
  1363--1379, 2015.

\bibitem{cohen2015distributed}
D.~Cohen, A.~Akiva, B.~Avraham, S.~Patterson, and Y.~C. Eldar, ``Distributed
  cooperative spectrum sensing from sub-nyquist samples for cognitive radios,''
  in \emph{2015 IEEE 16th International Workshop on Signal Processing Advances
  in Wireless Communications (SPAWC)}.\hskip 1em plus 0.5em minus 0.4em\relax
  IEEE, 2015, pp. 336--340.

\bibitem{sun2016collaborative}
Y.~Sun, B.~L. Mark, and Y.~Ephraim, ``Collaborative spectrum sensing via online
  estimation of hidden bivariate markov models,'' \emph{IEEE Transactions on
  Wireless Communications}, vol.~15, no.~8, pp. 5430--5439, 2016.

\bibitem{so2016group}
J.~So and W.~Sung, ``Group-based multibit cooperative spectrum sensing for
  cognitive radio networks,'' \emph{IEEE Transactions on Vehicular Technology},
  vol.~65, no.~12, pp. 10\,193--10\,198, 2016.

\bibitem{yao2017cluster}
F.~Yao, H.~Wu, Y.~Chen, Y.~Liu, and T.~Liang, ``Cluster-based collaborative
  spectrum sensing for energy harvesting cognitive wireless communication
  network,'' \emph{IEEE Access}, vol.~5, pp. 9266--9276, 2017.

\bibitem{goodfellow2016deep}
I.~Goodfellow, Y.~Bengio, and A.~Courville, \emph{Deep learning}.\hskip 1em
  plus 0.5em minus 0.4em\relax MIT press, 2016.

\bibitem{Steven}
S.~M. Kay, \emph{Fundamentals of Statistical Signal Processing}.\hskip 1em plus
  0.5em minus 0.4em\relax Prentice Hall PTR, 1993.

\bibitem{6247438}
S.~Kalyani and R.~M. Karthik, ``The asymptotic distribution of maxima of
  independent and identically distributed sums of correlated or non-identical
  gamma random variables and its applications,'' \emph{IEEE Transactions on
  Communications}, vol.~60, no.~9, pp. 2747--2758, September 2012.

\bibitem{8360549}
M.~{Srinivasan} and S.~{Kalyani}, ``Secrecy capacity of $\kappa-\mu$ shadowed
  fading channels,'' \emph{IEEE Communications Letters}, vol.~22, no.~8, pp.
  1728--1731, Aug 2018.

\bibitem{6957529}
S.~{Kumar}, G.~{Chandrasekaran}, and S.~{Kalyani}, ``Analysis of outage
  probability and capacity for$\kappa$-$\mu/\eta$-$\mu$faded channel,''
  \emph{IEEE Communications Letters}, vol.~19, no.~2, pp. 211--214, Feb 2015.

\bibitem{shaffer1995multiple}
J.~P. Shaffer, ``Multiple hypothesis testing,'' \emph{Annual review of
  psychology}, vol.~46, no.~1, pp. 561--584, 1995.

\bibitem{abdi2007bonferroni}
H.~Abdi, ``Bonferroni and {\v{s}}id{\'a}k corrections for multiple
  comparisons,'' \emph{Encyclopedia of measurement and statistics}, vol.~3, pp.
  103--107, 2007.

\bibitem{abdi2010holm}
------, ``Holm’s sequential bonferroni procedure,'' \emph{Encyclopedia of
  research design}, vol.~1, no.~8, pp. 1--8, 2010.

\bibitem{5937265}
P.~{Ray} and P.~K. {Varshney}, ``False discovery rate based sensor decision
  rules for the network-wide distributed detection problem,'' \emph{IEEE
  Transactions on Aerospace and Electronic Systems}, vol.~47, no.~3, pp.
  1785--1799, July 2011.

\bibitem{8452950}
V.~{Raj} and S.~{Kalyani}, ``Backpropagating through the air: Deep learning at
  physical layer without channel models,'' \emph{IEEE Communications Letters},
  vol.~22, no.~11, pp. 2278--2281, Nov 2018.

\bibitem{cybenko1989approximation}
G.~Cybenko, ``Approximation by superpositions of a sigmoidal function,''
  \emph{Mathematics of control, signals and systems}, vol.~2, no.~4, pp.
  303--314, 1989.

\bibitem{liu2019ensemble}
H.~Liu, X.~Zhu, and T.~Fujii, ``Ensemble deep learning based cooperative
  spectrum sensing with semi-soft stacking fusion center,'' in \emph{2019 IEEE
  Wireless Communications and Networking Conference (WCNC)}.\hskip 1em plus
  0.5em minus 0.4em\relax IEEE, 2019, pp. 1--6.

\bibitem{raj2017aggregating}
V.~Raj and S.~Kalyani, ``An aggregating strategy for shifting experts in
  discrete sequence prediction,'' \emph{arXiv preprint arXiv:1708.01744}, 2017.

\bibitem{7069496}
Z.~Yang, Y.~Song, and D.~Wang, ``An optimal operating frequency selection
  scheme in spectrum handoff for cognitive radio networks,'' in \emph{2015
  International Conference on Computing, Networking and Communications (ICNC)},
  Feb 2015.

\bibitem{5506438}
L.~Stabellini, ``Quantifying and modeling spectrum opportunities in a real
  wireless environment,'' in \emph{2010 IEEE Wireless Communication and
  Networking Conference}, April 2010, pp. 1--6.

\bibitem{sahoo2017online}
D.~Sahoo, Q.~Pham, J.~Lu, and S.~C. Hoi, ``Online deep learning: Learning deep
  neural networks on the fly,'' \emph{arXiv preprint arXiv:1711.03705}, 2017.

\end{thebibliography}
\end{document}